\documentclass[journal,article,accept,moreauthors,pdftex,preprints]{Definitions/mdpi} 
\usepackage{multirow}
\usepackage{amsmath}
\usepackage[version=4]{mhchem} 
\usepackage{savesym}
\savesymbol{tablenum}
\usepackage{siunitx}
\restoresymbol{SIX}{tablenum}
\usepackage{lscape}
\usepackage{longtable}
\usepackage{adjustbox,lipsum}

\firstpage{1} 
\pubvolume{xx}
\issuenum{1}
\articlenumber{5}
\pubyear{2019}
\copyrightyear{2019}
\history{Received: date; Accepted: date; Published: date}

\pdfoutput=1

\Title{The role of state-of-the-art quantum-chemical calculations in astrochemistry: formation route and spectroscopy of ethanimine as a paradigmatic case}

\Author{Carmen Baiano$^1$\orcidA{}, Jacopo Lupi$^1$\orcidB{}, Nicola Tasinato$^1$\orcidC{}, Cristina Puzzarini$^{2,*}$\orcidD{} and Vincenzo Barone$^{1,*}$\orcidE{}}

\address{
$^{1}$ \quad Scuola Normale Superiore, Piazza dei Cavalieri 7, I-56126 Pisa, Italy\\
$^{2}$ \quad Dipartimento di Chimica ``Giacomo Ciamician'', Universit\`a di Bologna, Via F. Selmi 2, 40126 Bologna, Italy}

\corres{Correspondence: cristina.puzzarini@unibo.it; vincenzo.barone@sns.it}

\abstract{The gas-phase formation and spectroscopic characteristics of ethanimine have been re-investigated as a paradigmatic case illustrating the accuracy of state-of-the-art quantum-chemical (QC) methodologies in the field of astrochemistry. According to our computations, the reaction between the amidogen, NH, and ethyl, \ce{C2H5}, radicals is very fast, close to the gas-kinetics limit. Although the main product channel under conditions typical of the interstellar medium leads to methanimine and the methyl radical, the predicted amount of the two E,Z stereoisomers of ethanimine is around 10$\%$. State-of-the-art QC and kinetic models lead to a [\ce{E{-}CH3CHNH}]/[\ce{Z{-}CH3CHNH}] ratio of ca. 1.4, slightly higher than the previous computations, but still far from the value determined from astronomical observations (ca. 3). An accurate computational characterization of the molecular structure, energetics, and spectroscopic properties of the E and Z isomers of ethanimine combined with millimeter-wave measurements up to 300 GHz, allows for predicting the rotational spectrum of both isomers up to 500 GHz, thus opening the way toward new astronomical observations. 
}


\begin{document}

\section{Introduction}

Until the latter part of the 20th century, it was believed that the vastness of interstellar space consisted mainly of hydrogen atoms. Apart from molecular hydrogen and a few other simple diatomic species, the harsh conditions of the interstellar medium (ISM) were thought to be incompatible with polyatomic molecules exhibiting even a small degree of complexity. This idea began to be questioned roughly fifty years ago with the discovery of the first ‘complex’ molecules (formaldehyde in 1969 \cite{snyder1969}, methanol in 1970 \citep{1970ApJ}, and formic acid in 1971 \cite{zuckerman1971}). Since these first hints, the pace of molecular detections in space has accelerated. The original paradigm alluded to above has now been completely erased by the discovery of more than two hundred molecular species in the ISM and circumstellar shells \cite{mcguire2018}.  
Molecules detected in space range from simple hydrides (such as \ce{H2}, \ce{H2O}, and \ce{NH}3), to hydrocarbon chain species (such as the cyanopolyynes, HC$_{2n+1}$N), to simple organics (such as alcohols and aldehydes), to highly reactive species such as ions and radicals (e.g. \ce{H2COH+}, \ce{HC3NH+}, and \ce{CH3O}), to polycyclic aromatic hydrocarbon molecules (PAHs) \cite{RevModPhys.85.1021}. The latter are the most abundant polyatomic species in space, although no individual PAH has yet been identified. This short and generic list makes it clear that interstellar chemistry is rich and diverse. Focusing on atomic elements, the contribution of hydrogen and helium amounts to about 98\%; therefore, that of heavier
elements (such as carbon, nitrogen, and oxygen) is only about 2\%. Nevertheless, as mentioned above and in the following, this small fraction of heavy elements makes possible a great variety of chemical compounds.

Among the molecules identified in the ISM, in the present context, the interest is on the so-called ``interstellar complex organic molecules'' (iCOMs), namely C-bearing molecules with at least six atoms \cite{herbst2009}, such as glycolaldehyde \cite{hollis2000} and acetamide \cite{hollis2006}. In this respect, molecular spectroscopy plays a crucial role in finding out whether a particular molecule is present in an astronomical environment and, if so, in what abundance (e.g. refs. \cite{jorgensen2012,zaleski2013,barone2015,emoca,ciccio,remoca}). Indeed, most of the gas-phase molecules mentioned above have been discovered thanks to radioastronomy, and thus by means of their rotational signatures. However, still unassigned features appear in radio-astronomical spectra, thus pointing out that we are far from a complete census of the interstellar molecules.
Even though iCOMs were first detected decades ago, the processes that lead to their production are still a matter of debate. While it is nowadays accepted that they can be synthesized either on dust-grain surfaces or by means of gas-phase chemistry \cite{Garrod06,Garrod08,Balucani15,Linnartz15,Vazart16,Rimola18}, the available literature reveals that the understanding of chemical evolution is at a rather primitive stage and current chemical models still show large deficiencies. In particular, in the last two decades, we have witnessed the growth of the grain-surface chemistry and the progressive abandonment of the gas-phase reactivity. However, recent detections of iCOMs in cold regions \cite{Vastel14, Balucani15}, where the surface chemistry cannot be efficient, have pointed out that gas-phase chemistry has been overlooked. The idea at the basis of grain-surface chemistry is that radical species trapped in icy mantles move and encounter a reaction partner, thus giving rise to rich chemistry \cite{Yamamoto}. This mechanism is called the Langmuir–Hinshelwood mechanism. A second possibility is offered by gaseous atoms and molecules that directly hit atoms and molecules on dust grains to form products. This is called the Rideal mechanism, which is --however-- less important in interstellar clouds \cite{Yamamoto}. It is thus evident that surface mobility is an important factor in grain-surface reactions. While at temperatures above 30 K all molecular species have a reasonable mobility, at lower temperatures the mechanisms described above cannot be efficient (in particular, at T = 10 K only hydrogen atoms have a non-negligible mobility). 

On the one hand, the detection of iCOMs showing a certain degree of complexity reveals an undisputed molecular complexity in the universe, which is --however-- not entirely disclosed. On the other hand, there is still much to be understood about how iCOMs, and more generally molecules, can be formed in the harsh conditions typical of the ISM and how they can evolve toward more complex species. It is thus clear that we are facing two major challenges in astrochemistry, namely the detection of `new' iCOMs and the understanding of chemical reactivity, and that they are strongly interconnected \cite{fspas.2020.00019}. For this reason, we present an integrated experimental and theoretical strategy that address both challenges \cite{D0CP00561D}: state-of-the-art computational approaches are combined with laboratory spectroscopy. Ethanimine, a prebiotic iCOM (as will be explained later), has been chosen as case study to illustrate this strategy.    

Computational chemistry offers an invaluable aid in providing two main pieces of information, namely: \\
(1) The investigation of reactive potential energy surfaces (PESs) from both energetic (thermochemistry) and kinetic points of view. Two possibilities can be actually envisaged: (i) starting from purposely chosen precursors the formation route of the sought product (i.e. a molecule already identified in the ISM) is derived; (ii) starting from small reactive species the possible pathways are elaborated. In both cases, the harsh conditions of the ISM (extremely cold --down to 10 K-- regions, extremely low density --even 10$^4$ particles/cm$^3$) are used as constraints.   \\
(2) Referring to point (1.ii), the accurate prediction of the spectroscopic parameters of those products that can be of interest and for which such information is still missing is carried out. \\
The second piece of information (point (2)) is then used to guide and support laboratory measurements that, in the field of rotational spectroscopy, are mandatory to further proceed toward astronomical searches. Finally, astronomical data can confirm the plausibility of the developed formation pathway. For example, the identification (in astronomical surveys) of the sought product of the reaction investigated can provide an indirect confirmation. Astronomical observations can also be used to support the effectiveness of a gas-phase formation route \cite{formamide-solis}. Furthermore, as in the case of ethanimine, i.e. in the presence of two different isomers, the computed branching ratios can be compared with the relative astronomical abundance.      

As briefly mentioned above, ethanimine has been chosen as a prototypical molecular system for presenting and discussing our strategy. Its astrophysical relevance is related to its prebiotic potential as a possible precursor of amino acids \cite{quan2016}, like alanine, by reaction with HCN and \ce{H2O}, or with formic acid (\cite{woon2002,elsila2007,loomis2013}). Both isomers of ethanimine have been detected in Sagittarius B2 North, SgrB2 (N), as a result of the GBT-PRIMOS project (this acronym standing for Green Bank Telescope - PRebiotic Interstellar MOlecule Survey) \cite{loomis2013}. Furthermore, this molecule can also be of interest in the study of the nitrogen-based Titan's atmosphere \cite{balucani2010}. As is well known, Titan is considered a good approximation of the primitive Earth \cite{Clarke97,Raulin09}, thus explaining its relevance in the study of prebiotic chemistry in space.

\subsection{The Thermochemistry-kinetics-spectroscopy Strategy}
 
As explained above, our strategy can start from a potential iCOM or from a potential gas phase reaction between two molecules already detected in the ISM. The example addressed in this contribution belongs to the first case. In any event, the starting point is the design of a feasible and accessible reactive PES leading to the species of interest in the first case or starting from the chosen reactants in the latter option. This preliminary investigation is performed at a relatively low computational level, relying on density functional theory (DFT). Usually, different pathways can be derived and only those that are feasible in the conditions typical of the astronomical environment under consideration are further investigated at a higher level \cite{Salta2000}. Therefore, the more promising pathways are re-investigated in order to improve the structural determination of the stationary points as well as to compute the energetics at the state of the art. To conclude, kinetic calculations are carried out, these providing the conclusive information on the feasibility of the suggested mechanisms and on the branching ratios of the reaction products \cite{Barone2015b,0004-637X-854-2-135}. On the spectroscopic side, the target species are better characterized in terms of structural, molecular and spectroscopic properties in order to provide reliable predictions for the laboratory spectroscopy studies \cite{Puzzarini2019,D0CP00561D}. Finally, astronomical observations can be performed and might be used to support the outcomes of the theoretical investigation (see e.g. Ref. \cite{Skouteris2017,0004-637X-854-2-135}).

In the case of ethanimine, several pieces of information and reliable data were already available prior to our investigation. This was indeed the reason why it was chosen as case study to develop and test our strategy. Ethanimine was already detected in the ISM, but the knowledge on its rotational spectrum was limited. Possible gas-phase mechanisms were also available in the literature, even if not entirely satisfactory from an astronomical point of view.

\subsubsection{Formation pathway}

As far as the formation mechanism of ethanimine in the ISM is concerned, among the gas-phase reactions considered by Quan \emph{et al.} \cite{quan2016}, the \ce{NH + C2H5} reaction is certainly the best candidate, because the analogous \ce{NH + CH3} reaction was demonstrated to be the dominant formation route of methanimine in the model proposed by Suzuki \emph{et al.} \cite{suzuki2016}.
Recently, Balucani and coworkers \cite{balucani2018} performed an exhaustive search of different possible reaction paths, concluding that only the following three channels are open in the ISM conditions:

\begin{equation}
\ce{NH + C2H5 -> CH3 + CH2NH} \tag{R1}
\end{equation}
\begin{equation}
\ce{NH + C2H5 -> E{-}CH3CHNH + H} \tag{R2a}
\end{equation}
\begin{equation}
\ce{NH + C2H5 -> Z{-}CH3CHNH + H} \tag{R2b}
\end{equation}
While the computational level of ref. \cite{balucani2018} might not be sufficient to obtain quantitative results, the computed energy barriers governing the other considered paths are so high that they can be safely excluded in the present study, with the only exception of channel 3: 

\begin{equation}
\ce{NH + C2H5 -> C2H4 + NH2} \tag{R3}
\end{equation}

The reaction of \ce{CH2} with \ce{CH2NH} was also proposed \cite{singh2018}. However, the computational level employed to analyze the mechanism was quite low and only the singlet form of \ce{CH2} leads to submerged barriers, thus requiring a spin-orbit coupling mechanism from the more stable triplet form. Furthermore, the first step of this reaction is the abstraction of one hydrogen atom from \ce{CH2NH} by \ce{CH2}. Therefore, the effective reaction involves the \ce{CH3} and CHNH radicals:

\begin{equation}
\ce{CH3 + CHNH -> E{-}/Z{-}CH3CHNH} \tag{R4a/4b}
\end{equation}
While the \ce{CH3} radical has been detected in several regions of the ISM, only the \ce{CHNH+} cation has been identified so far, the corresponding neutral radical remaining still unobserved. As a consequence we have not investigated in detail this alternative mechanism.

The rate coefficients employed by Quan \emph{et al.} \cite{quan2016} in their model are 8.25 \AA $\times 10^{-12}$ cm$^3$/s for \ce{NH + C2H5} $\rightarrow$ \ce{E{-}CH3CHNH + H} (R2a) and 2.75 \AA $\times 10^{-12}$ cm$^3$/s for \ce{NH + C2H5} $\rightarrow$ \ce{Z{-}CH3CHNH + H} (R2b) with no temperature dependence. However, in their model, they did not take into consideration that the channel (R1) consumes a fraction of the reactants, thus reducing the total production rate of ethanimine. The lack of temperature dependence being quite surprising, the computations of ref. \cite{balucani2018} indeed pointed out that the rate constants for channels R2a and R2b are very similar to one another and both of them nearly double when increasing the temperature from 10 K to 200 K (i.e. the interval considered in ref. \cite{quan2016}).

\subsubsection{Spectroscopic Characterization}

As clear from the preceding section and the literature cited, ethanimine exists in two isomeric forms, E and Z (see Figure \ref{fig:specie}), with the former being the most stable. As already mentioned, both isomers have been detected in the ISM \cite{loomis2013}. However, the experimental works on ethanimine rotational spectrum that allowed such an identification were limited to low frequencies, i.e. below 140 GHz \cite{brown1980,lovas1980,loomis2013}. Since extrapolations from low-frequency laboratory measurements usually provide inaccurate predictions for higher frequencies and, in view of the extended astronomical observatory facilities provided by the Atacama Large Millimeter/submillimeter Array (ALMA; working frequency range: 84-950 GHz), the extension of the investigation of the rotational spectra for both ethanimine isomers was warranted.
To improve the spectroscopic characterization of both isomeric forms of \ce{CH3CHNH}, an accurate computational study of the energetics as well as structural and spectroscopic parameters was performed \cite{melli2018}, thereby focusing on the description of the methyl internal rotation and on the centrifugal distortion terms. 

In ref. \cite{melli2018}, an accurate characterization of the vibrational (infrared, IR) spectrum was also provided. Indeed, in view of predicting the gas-phase IR spectrum, the computed fundamental frequencies of ref. \cite{melli2018} can be considered more accurate and reliable than those from the low-resolution gas-phase study of ref. \cite{hashiguchi1984} and the matrix investigation reported in ref. \cite{stolkin1977}. In fact, both studies provided either incomplete or contradictory results. Therefore, the computational work on the IR spectra of both isomers of ethanimine might be of great help in future observation of planetary atmospheres (e.g. that of Titan) relying on, for instance, the James Webb Space Telescope to be launched in 2021.

\subsection{Organization of the Paper}

In the following section the computational methodology at the basis of both challenges, namely the investigation of the formation pathway and the spectroscopic characterization, will be described in detail. On these grounds, the purpose of the present paper is to provide reliable and accurate rate constants for the reaction mechanisms considered by exploiting state-of-the-art quantum-chemical (QC) and kinetic computations. At the same time, we reconsider the spectroscopic parameters of ethanimine, which were previously reported in ref. \cite{melli2018}. 
In the subsequent section, the thermochemical, kinetic and spectroscopic results will be presented and discussed. Finally, concluding remarks will address the effectiveness of our strategy.

\section{Computational Methodology}

Our QC strategy relies on a composite scheme in order to obtain high accuracy at an affordable computational cost. This combines coupled-cluster (CC) techniques with extrapolation to the complete basis set (CBS) limit and consideration of core-valence (CV) correlation contributions. To account for vibrational effects, also including anharmonicity, we resort to second-order vibrational perturbation theory (VPT2) \cite{Barone2005}. For the evaluation of kinetic parameters, a master equation (ME) approach based on ab initio transition state theory (AITSTME) \cite{Klippenstein2017} is used.

\subsection{Electronic Structure calculations}\label{sec:EleStruc}

The QC calculations described in the following have been performed with Gaussian 16 \cite{g16} and CFOUR \cite{cfour} quantum-chemical packages. The former code has been employed for DFT and VPT2 computations, while the latter for CC calculations. Furthermore, the MRCC code \citep{mrcc} interfaced to CFOUR has been used to perform the CC calculations including quadruple excitations.

\subsubsection{Reactive potential energy surface}\label{r-pes}

A preliminary characterization of the reactive PES has been carried out at the B3LYP-D3/aug-cc-pVTZ level \cite{LeeYangParr1988,Becke1993,kendall1992a,Dunning1989} (hereafter shortly denoted as B3), with geometry optimizations combined with harmonic force-field calculations in order to characterize the stationary points. More generally, for this introductory analysis of the reactive PES, a double-zeta quality basis set can be safely employed in conjunction with the hybrid B3LYP functional. 
The results obtained at the B3 level were found already in agreement with those reported in literature \cite{balucani2018}. 

\begin{figure}[H]
\centering
\includegraphics[width=0.8\textwidth]{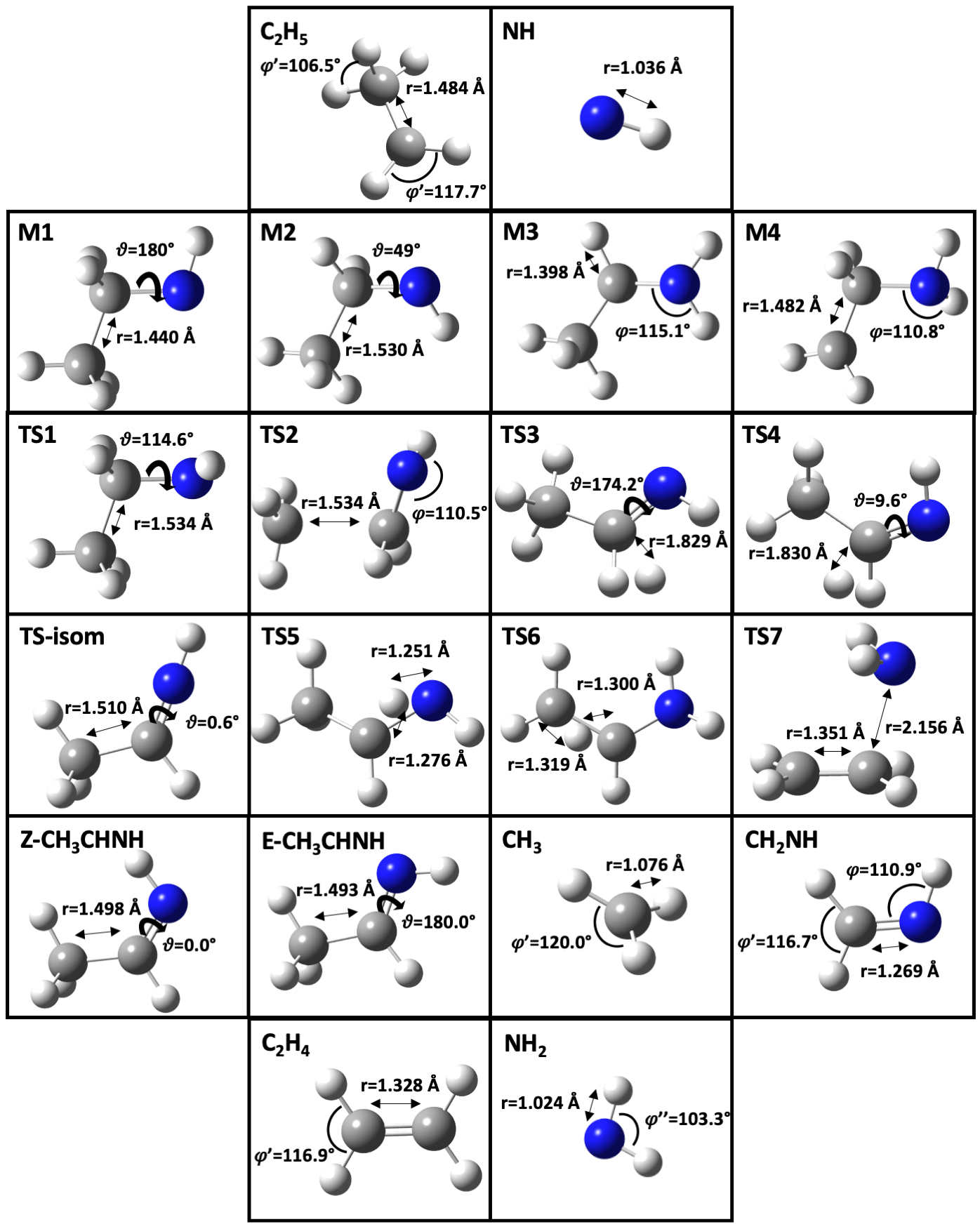}
\caption{Structures and selected geometrical parameters of all stationary points on the ethanimine PES optimized at the B2 level. Bond lengths are in \SI{}{\angstrom}, whereas HNC ($\varphi$), HCH ($\varphi$') or HNH ($\varphi$'') valence angles and CCNH ($\vartheta$) dihedral angles are in degrees.} \label{fig:specie}
\end{figure} 

Since the double-hybrid B2PLYP functional \cite{Grimme2006} represents a good compromise between accuracy and computational cost for both structural and spectroscopic properties \cite{C9CP06768J,anie.201906977,Boussessi2000} and is able to provide a reliable description of reactive and non-reactive PESs \cite{Barone2015b,0004-637X-854-2-135,Spada2017a,Spada2017b,anie.201906977,C9CP06768J,Lupi2020,Puzzarini2020}, our strategy relies on it in order to provide the final description of the stationary points from a structural point of view. For this purpose, B2PLYP is usually employed in conjunction with a triple-zeta basis sets incorporating diffuse functions. In the present case, the aug-cc-pV$n$Z ($n$=T,Q) basis sets have been employed. In addition, different basis sets have been tested. However, these as well as the aug-cc-VQZ set have not pointed out any substantial modification of either the PES or geometrical parameters with respect to the combination of B2PLYP and aug-cc-pVTZ. Furthermore, to account for dispersion interactions, the DFT-D3 scheme \cite{GrimmeD3} has been used coupled to the Becke-Johnson (BJ) damping function \cite{Grimme2011}. Overall, in the following as well as in Figure \ref{fig:specie}, we only refer to the B2PLYP-D3(BJ)/aug-cc-pVTZ level of theory, hereafter simply denoted as B2. The interested reader can find a detailed structural analysis in the Supplementary Information (SI) material. In passing, it should be noted that the NH species has been considered only in the most stable triplet state ($^{3}\Sigma^{-}$) \cite{Rajvanshi2010}.

The subsequent step of our strategy requires the accurate estimates of reaction energies and barriers. For this purpose, we rely on composite schemes based on coupled-cluster (CC) theory and exploited on top of B2 optimized geometries. In particular, two different types of approach have been employed: (i) schemes entirely based on CC calculations (denotes as ``CBS+CV'', ``CBS+CV+fT+pQ'', and ``HEAT-like''), and (ii) the so-called ``cheap'' model, which involves an important reduction of the computational cost, while keeping high accuracy. All these approaches are detailed in Section \ref{sec:extr}. 

Finally, the inclusion of zero-point vibrational energies (ZPVEs) is required. These can be computed both at harmonic and anharmonic levels. In this respect, as briefly mentioned above, the nature of the stationary points has been checked by the evaluation of analytical Hessians (harmonic force fields), which --in turn-- provide harmonic ZPVEs. Anharmonic force fields have also been calculated at the B2 level, thus allowing the evaluation of anharmonic ZPVEs by means of the generalized vibrational-perturbation theory (GVPT2). With respect to standard VPT2, GVPT2 excludes resonant terms from the perturbative treatment and reintroduces them through a reduced-dimensionality variational calculation \cite{Martin1995,Barone2004,Barone2005,Bloino2012,Puzzarini2019}.

\subsubsection{Spectroscopic parameters}\label{sec:spec-det}

As mentioned in the Introduction, most of the molecular species detected in the ISM have been identified thanks to their rotational fingerprints. Indeed, the rotational spectrum of a given molecule is so characteristic that the observation of its rotational transitions provides an unequivocal proof of the presence of that species in the environment under consideration. For this reason, rotational spectroscopy is the spectroscopic technique taken into consideration in our strategy.

The leading terms of rotational spectroscopy are the rotational constants, the rotational transition frequencies mainly depending on their values \cite{Gor84}. In general, to provide accurate predictions of rotational transitions, it is mandatory to go beyond the rigid-rotor approximation, thus accounting for centrifugal-distortion and vibrational effects. To incorporate the latter in the prediction of rotational constants, vibrational corrections ($\Delta B_{vib}$) are added to the equilibrium rotational constants ($B_e$). Within VPT2, the vibrational ground-state rotational constants ($B_0$) are given by the expression \cite{Mills72}:

\begin{equation}
B_0^i = B_e^i + \Delta B_{vib}^i = B_e^i - \frac{1}{2} \sum_r \alpha_r^i \; .
\label{se}
\end{equation}
with $i$ denoting the inertial axis ($i$ = $a$, $b$, $c$; that is, for instance, $B_0^a$ = $A_0$). The summation runs over all vibrational normal modes and the $\alpha_r^i$'s are the vibration-rotation interaction constants. 

It is important to note that the vibrational corrections provide a small contribution to $B_0$'s, indeed accounting only for $\sim$1-3\% of the total value. However, as pointed out in refs. \cite{rotstat,benchS}, the inclusion of vibrational corrections is fundamental for quantitative predictions.
From a computational point of view, the equilibrium rotational constants are straightforwardly derived from the equilibrium structure, while vibrational corrections require anharmonic force-field calculations \cite{Puzzarini-IRPC2010_Review_rotational_spc}. In view of the fact that $B_e$ provides the largest contribution to $B_0$, the major computational effort is put on geometry optimizations. To fulfil the accuracy requirements, composite schemes are thus employed. In particular, the CBS+CV approach mentioned above (detailed in Section \ref{sec:extr}) has been chosen. Vibrational corrections have been determined at the B2 level (however employing a smaller basis set, namely a modified version of maug-cc-pVTZ \cite{Pap09,C5CP07386C}). 

To account for the effect of centrifugal distortion on rotational transitions, the computation of the corresponding parameters is required. 
The quartic centrifugal-distortion constants are obtained as a by product of harmonic force-field computations \cite{Puzzarini-IRPC2010_Review_rotational_spc}, 
while the sextic ones are derived from the knowledge of the cubic force field. Therefore, the latter terms have been evaluated at the B2 level. In ref. \cite{melli2018}, the harmonic force field has been computed at a higher level of theory, thereby employing the CCSD(T) method in conjunction with the cc-pCVQZ basis set, with all electrons being correlated (all).

\begin{figure}[H]
\centering
\includegraphics[width=0.6\textwidth]{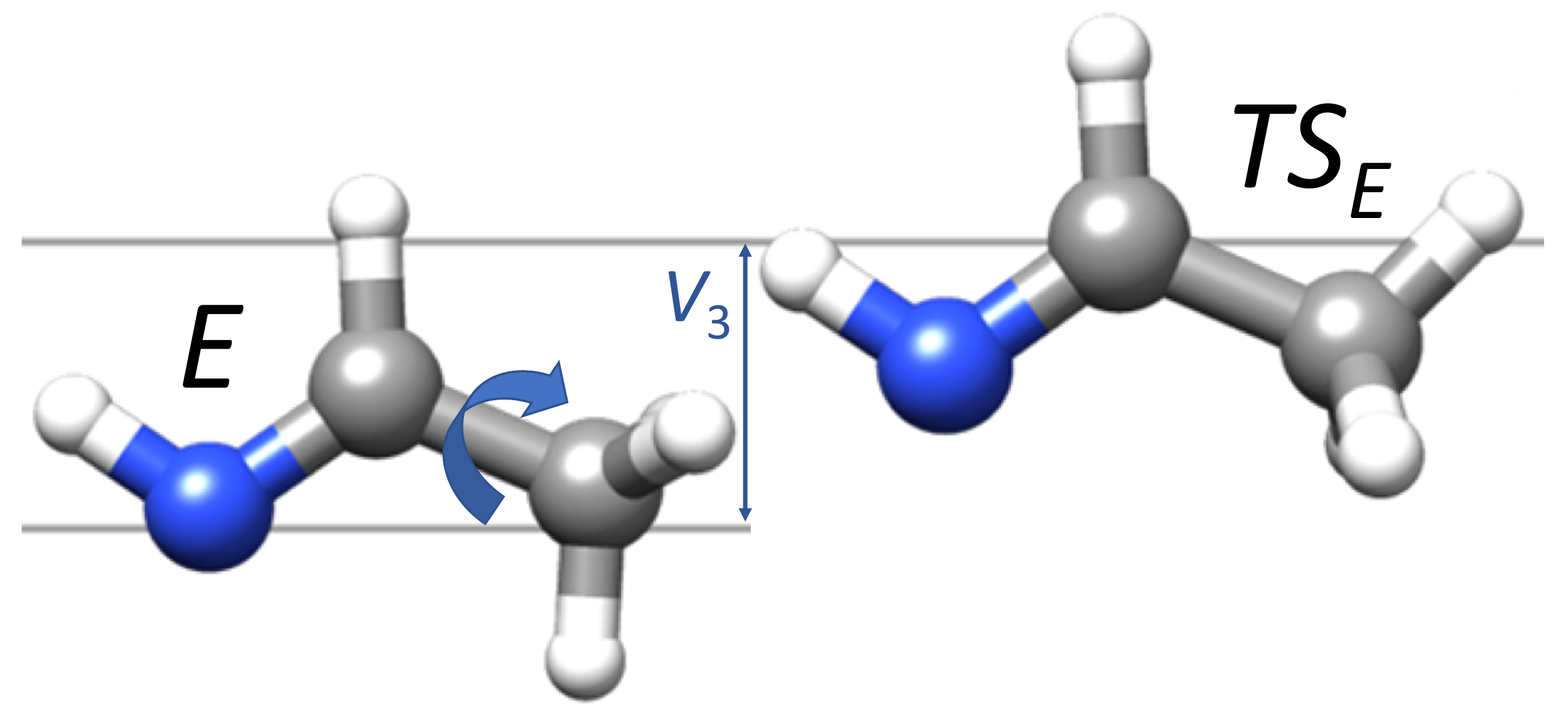}
\caption{Methyl internal rotation in E-ethanimine} \label{fig:v3}
\end{figure}

In the specific case of ethanimine, to complete the characterization of the rotational spectrum, nitrogen quadrupole coupling constants and the spectroscopic parameters related to the methyl internal rotation are needed. To evaluate the former constants, the electric field gradient (EFG) tensor is required and it has been evaluated, for both isomers, at the all-CCSD(T)/cc-pCVQZ level using the CBS+CV reference geometry. These have been further augmented by vibrational corrections at the B2 level. For all details related to the conversion of the EFG elements to the nuclear quadrupole coupling constants, interested readers are --for example-- referred to ref. \cite{Puzzarini-IRPC2010_Review_rotational_spc}. To model the methyl internal rotation, the barrier for the hindered rotation has been derived as energy difference between the E-/Z-\ce{CH3CHNH} minima and the corresponding transition state (see Figure \ref{fig:v3} and ref. \cite{melli2018} for details).

\subsubsection{Composite schemes}\label{sec:extr}

The composite schemes described in the following rely on or start from the CCSD(T) method, the acronym standing for single and double excitations and a perturbative treatment of triple excitations \cite{Purvis1982}. The coupled cluster $\mathcal{T}_1$ diagnostic \cite{qua.560360824} has been found to be smaller than 0.02 for closed-shell molecules and smaller than 0.04 for open-shell species (see Table 3 in the SI), thus confirming that the non-dynamical correlation is negligible for all the systems investigated.

\paragraph{The CC-based approaches}
The composite scheme denoted as CBS+CV accounts for the extrapolation to the CBS limit and CV corrections at the CCSD(T) level. Within this approach, electronic energies are obtained as follows:

\begin{equation}
    E_{\mathrm{CBS+CV}}=E^{HF-SCF}_{\mathrm{\infty}} + \Delta E^{corr}_{\mathrm{\infty}} + \Delta E(CV) \; . 
\label{cbscv}	   
\end{equation}
The terms on the right-hand side correspond, respectively, to: (1) the Hartree-Fock (HF-SCF) electronic energy extrapolated to the CBS limit using the exponential form proposed by Feller \cite{Feller1993} combined with the cc-pV$n$Z basis sets \cite{Dunning1989} ($n$=T,Q,5); (2) the frozen-core (fc) CCSD(T) correlation energy extrapolated to the CBS limit employing the two-point $n^{-3}$ formula proposed by Helgaker \emph{et al.} \cite{Helgaker1997} in conjunction with the cc-pV$n$Z sets, with either the $n$=T,Q or $n$=Q,5 combination of bases; (3) the CV correlation correction is calculated as energy difference between all and fc CCSD(T), both with the same basis set. The cc-pCVTZ and cc-pCVQZ \cite{Woon&Dunning1995} bases have been employed depending on whether the extrapolation to the CBS limit has been done using the $n$=T,Q or $n$=Q,5 combinations of basis sets, respectively. 

To further improve the energetics, a HEAT-like approach can be (and indeed has been) exploited, the reference of this scheme being the HEAT protocol \cite{StantonHEAT-2004}. The latter is known to provide sub-kJ mol$^{-1}$ accuracy. The HEAT-like model is obtained by incorporating additional terms into equation \ref{cbscv}:

\begin{equation}
E_{tot} = E_{\mathrm{CBS+CV}}  + \Delta E_{\mathrm{fT}}
	   + \Delta E_{\mathrm{pQ}} + \Delta E_{\mathrm{REL}} + \Delta E_{\mathrm{DBOC}}  \; ,
\label{heat}	   
\end{equation}
where $E_{\mathrm{CBS+CV}}$ is the CBS+CV energy (equation \ref{cbscv}). Similarly to the CV contribution, corrections due to the full treatment of triple excitations, $\Delta E_{\mathrm{fT}}$, and the perturbative description of quadruples, $\Delta E_{\mathrm{pQ}}$, are computed as energy differences between CCSDT \cite{ccsdt1,ccsdt2,ccsdt3} and CCSD(T) and between CCSDT(Q) \cite{1.1950567,1.2121589,1.2988052} and CCSDT (all of them within the fc approximation) employing the cc-pVTZ and cc-pVDZ basis sets, respectively. Inclusion of fT and pQ contributions leads to the definition of the CBS+CV+fT+pQ scheme. The diagonal Born-Oppenheimer correction \cite{dboc1,dboc2,dboc3,dboc4}, $\Delta E_{\mathrm{DBOC}}$, and the scalar relativistic contribution to the energy, $\Delta E_{\mathrm{REL}}$,\cite{rel1,rel2} are also included in the HEAT-like approach. The former corrections are usually computed at the HF-SCF/aug-cc-pV$n$Z level \cite{kendall1992a}, with $n$=D,T, and the relativistic corrections are obtained at the all-CCSD(T)/aug-cc-pCV$n$Z level ($n$=D,T) and include the (one-electron) Darwin and mass-velocity terms. In the present study, the last two corrections have been obtained using double-zeta basis sets, even if their convergence with respect to contributions calculated with triple-zeta basis sets has been checked for a few stationary points.

To exploit the high accuracy of the CBS+CV energy approach (equation \ref{cbscv}) for molecular structure determinations, the energy-gradient composite scheme introduced in refs.~\cite{Miriam2,Miriam1} can be employed. The resulting energy gradient to be employed in geometry optimizations is given by the following expression:

\begin{equation}
\label{best-geo} 
\frac{dE_{\mathrm{CBS+CV}}}{dx}  =
\frac{dE^{HF-SCF}_{\mathrm{\infty}}}{dx} + \frac{d\Delta E^{corr}_{\mathrm{\infty}}}{dx}  +  \frac{d\Delta E(\mathrm{CV})}{dx} \; ,
\end{equation}
where the terms on the right-hand side are defined as in equation \ref{cbscv}. Analogously to equation \ref{heat}, the energy gradient of equation \ref{best-geo} can be further improved by including the contributions due to full treatment of triples and quadruple excitations.

\paragraph{The ChS approach}
In the present work, electronic energies have also been computed by resorting to a computationally less demanding strategy, the so-called ``cheap'' scheme (hereafter denoted as ChS). Our strategy relies on this model when larger systems are involved. In the present context, it has been employed to further demonstrate its high accuracy and reliability, which was already pointed out in refs. \cite{D0CP00561D,Salta2000,Lupi2020}.

This approach has been originally developed for the determination of accurate structural parameters \cite{Puzzarini2011,Puzzarini2013} and then extended to energy evaluations \cite{Puzzarini2014}. In this scheme, the fc-CCSD(T)/cc-pVTZ energy is taken as the starting point and augmented by contributions accounting for the extrapolation to the CBS limit and CV correlation. To keep the computational cost limited, contrary to the CBS+CV approach, these terms are computed using second-order {M{\o}ller}-Plesset perturbation theory (MP2) \cite{PhysRev.46.618}:

\begin{equation}
E_{ChS}=E^{CCSD(T)/VTZ} +\Delta E^{MP2/CBS} + \Delta E^{MP2/CV}  \; 
\label{cheap}
\end{equation}
The second term on the right-hand side is the contribution due to the extrapolation to the CBS limit, which is performed in two steps (i.e. by separating the HF-SCF and correlation contributions) as in the CBS+CV scheme, and employing the same extrapolative formulas. MP2 calculations are carried out in conjunction with the cc-pVTZ and cc-pVQZ basis sets. The third term is evaluated, as in equation \ref{cbscv}, as `all-fc' energy difference at the MP2/cc-pCVTZ level.

\subsection{Kinetic Models}\label{sec:kinmod}

Phenomenological rate coefficients have been obtained within the AITSTME approach using the Rice-Ramsperger-Kassel-Marcus (RRKM) 1D master equation system solver (MESS) code \cite{georgievskii2013reformulation}. For channels possessing a distinct saddle point, rate coefficients have been determined by conventional transition state theory (TST) within the rigid-rotor harmonic-oscillator (RRHO) approximation. Instead, rate constants for barrierless elementary reactions have been computed employing phase space theory (PST) \cite{pechukas1965detailed,chesnavich1986multiple}, again within the RRHO assumption. 

For what concerns PST, it provides a useful, and easily implemented, reference theory for barrierless reactions, and it largely used in computational kinetics applied to astrochemistry (see, e.g., refs. \cite{0004-637X-810-2-111,Skouteris2017,10.1093/mnras/sty2903,0004-637X-854-2-135}). Its assumption is that the interaction between the two reacting fragments is isotropic and does not affect the internal fragment motions. This approximation is especially valid if the dynamical bottleneck lies at large separations where the interacting fragments have free rotations and unperturbed vibrations. This is generally true for low temperature phenomena, as is the case for the ISM. The isotropic attractive potential is assumed to be described by the functional form $-\frac{C}{R^6}$, where the coefficient $C$ is obtained by fitting the electronic energies obtained at various long-range distances of fragments. To be more precise, we performed a relaxed scan of the \ce{HN{-}CH2CH3} distance and then fitted the corresponding minimum energy path to a $f(x)=f_0-\frac{C}{x^6}$  function, thus obtaining a $C$ value of \SI{118.86}{\bohr\tothe{6}\hartree}. Tunnelling has been take into account by using the Eckart model \cite{eckart1930penetration}.

\begin{figure}[H]
\centering
\includegraphics[width=1.05\textwidth]{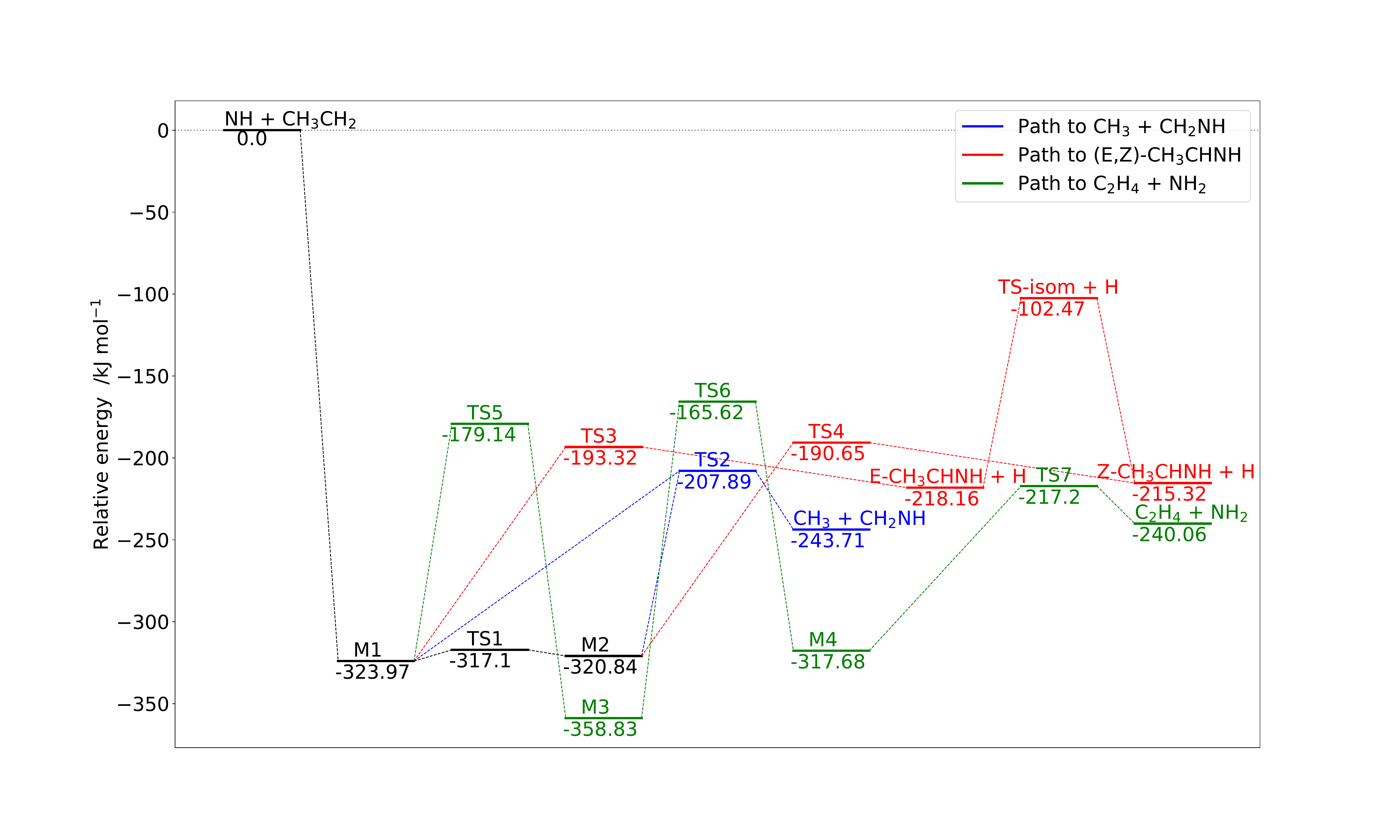}
\caption{Reaction mechanisms for the \ce{NH + C2H5} reaction: CBS+CV energies augmented by B2 anharmonic ZPVEs.}\label{fig:pes}
\end{figure}

\section{Results and Discussion}

The thermochemical aspects of the gas-phase reaction between the ethyl and imidogen radicals are first discussed, and then the attention will be focused on the results of RRKM kinetic simulations. Subsequently, the spectroscopic characterization of ethanimine is addressed.

\subsection{The \texorpdfstring{NH + C$_2$H$_5$}{NH + C2H5} reaction: thermochemistry}

As described in the Introduction, we have characterized the most favorable reaction channels following the addition of \ce{NH} to \ce{C2H5}, namely R1, R2a, R2b and R3. The resulting pathways are presented in Figure \ref{fig:pes}.

The initial association of reactants leading to the M1 intermediate is highly exothermic, with M1 easily interconverting to M2 through a small energy barrier (TS1). 
Since the channel leading to \ce{CH3 + CH2NH} requires the smallest number of steps and the lowest barrier (TS2; breaking of the \ce{C\bond{-}C} bond), thermodynamic considerations should favor this process. The evolution of M1 to \ce{E{-}CH2CHNH + H} and of M2 to \ce{Z{-}CH2CHNH + H} requires overcoming the transition states TS3 and TS4, respectively, which imply the breaking of one \ce{C\bond{-}H} bond and are less than \SI{10}{\kilo\joule\per\mol} higher in energy than TS2. This means that multiple competitive pathways can come into play. Since the \ce{E <-> Z} isomerization involves a high energy transition state, i.e. TS-isom, once ethanmine is formed, the direct interconversion of the two isomers  in the gas phase is very unlikely. As expected, the predicted stability of \ce{E{-}CH3CHNH} by \SI{2.84}{\kilo\joule\per\mol} with respect to the Z isomer is in agreement with the value of \SI{2.77}{\kilo\joule\per\mol} by Melli \emph{et al.} \cite{melli2018}, the level of theory being nearly the same.
Analogously, the \ce{E <--> Z} isomerization barrier of \SI{115.7}{\kilo\joule\per\mol} (calculated with respect to \ce{E{-}CH2CHNH}) matches that of ref. \cite{melli2018}, while the CCSD(T)/aug-cc-pVTZ value reported in ref. \cite{balucani2018} results to be overestimated.
Besides leading to methanimine or ethanimine, M1 could potentially undergo multiple hydrogen transfers, according to the path reported in green in Figure \ref{fig:pes}, to end up with the release of \ce{NH2} and \ce{C2H4}. However, the multiple rearrangements and the relatively high barrier associated to the M3 - M4 conversion make this process the less feasible in the harsh conditions of the ISM.

The electronic energies of all the species involved in the paths shown in Figure \ref{fig:pes}, referred to the isolated reactants, are collected in Table \ref{tab:energie}, where harmonic and anharmonic ZPVE corrections are also given. In addition to CBS+CV, CBS+CV+fT+pQ, HEAT-like and ChS energies, Table \ref{tab:energie} also reports B2 results and, for the sake of comparison, the results from Ref. \cite{balucani2018}.

\begin{table}[t!]
\centering
\caption{Relative electronic energies together with ZPVE corrections. Values in \SI{}{\kilo\joule\per\mol}.}
\begin{adjustbox}{max width=\textwidth}
\small
\begin{tabular}{@{}cccccccccc@{}}
\toprule
  & B2 & CBS+CV\textsuperscript{\emph{a}}& CBS+CV+fT+pQ\textsuperscript{\emph{b}}& HEAT-like\textsuperscript{\emph{b}} & ChS  &  harm-ZPE\textsuperscript{\emph{c}}     &   anharm-ZPE\textsuperscript{\emph{c}} &$\Delta H^0_0$\textsuperscript{\emph{d}} &$\Delta H^0_0$\textsuperscript{\emph{e}}    \\ \midrule
\ce{NH + C2H5}	      & 0.00    & 0.00      & 0.00     &    0.00   & 0.00    & 0.00  & 0.00 & 0.00   & 0.00    \\            
M1                    & -345.16 & -352.10   & -351.92  & -352.09   & -352.82 & 28.40 & 28.13&-323.97 & -311    \\            
                      &         & (-352.01) & (-351.83)& (-352.00) &         &       &      &        & (-323)  \\            
M2                    & -342.24 & -349.43   & -349.27  & -349.44   & -350.28 & 28.94 & 28.59&-320.84 & -308    \\            
                      &         & (-349.33) & (-349.18)& (-349.35) &         &       &      &        & (-320)  \\            
M3                    & -382.90 & -388.91   & -        & -         & -390.12 & 31.07 & 30.08&-358.83 & -340    \\            
                      &         & (-388.89) &          &           &         &       &      &        & (-357)  \\            
M4                    & -336.75 & -345.98   & -        & -         & -347.39 & 28.74 & 28.30&-317.68 & -301    \\            
                      &         & (-345.71) &          &           &         &       &      &        & -       \\            
TS1                   & -338.10 & -345.27   & -345.11  & -345.28   & -346.14 & 28.51 & 28.17&-317.10 & -304    \\            
                      &         & (-345.20) & (-345.04)& (-345.21) &         &       &      &        & (-316)  \\            
TS2                   & -223.85 & -226.32   & -229.11  & -229.22   & -227.56 & 18.47 & 18.43&-207.89 & -197    \\            
                      &         & (-226.58) & (-229.37)& (-229.48) &         &       &      &        &  (-211) \\            
TS3                   & -195.51 & -202.17   & -203.93  & -204.09   & -202.88 & 9.31  & 8.85 &-193.32 & -180    \\            
                      &         & (-202.41) & (-204.18)& (-204.34) &         &       &      &        & (-197)  \\            
TS4                   & -192.92 & -199.72   & -201.47  & -201.62   & -200.42 & 9.48  & 9.07 &-190.65 & -177    \\            
                      &         & (-199.95) & (-201.70)& (-201.85) &         &       &      &        & (-194)  \\            
TS-isom               & -94.55  & -98.47    & -98.17   & -98.40    & -100.55 & -4.13 & -4.00&-102.47 & -84     \\            
                      &         & (-98.81)  & (-98.51) & (-98.74)  &         &       &      &        & (-101)  \\            
TS5                   & -190.43 & -197.33   & -        & -         & -199.01 & 18.79 & 18.19&-179.14 & -165    \\            
                      &         & (-197.27) &          &           &         &       &      &        &  -      \\            
TS6                   & -176.42 & -182.49   & -        & -         & -184.40 & 17.70 & 16.87&-165.62 & -146    \\            
                      &         & (-182.51) &          &           &         &       &      &        &  -      \\            
TS7                   & -236.01 & -237.83   & -        & -         & -238.77 & 20.67 & 20.63&-217.20 & -207    \\            
                      &         & (-238.27) &          &           &         &       &      &        &  -      \\            
\ce{CH2NH + CH3}      & -245.85 & -251.18   & -251.36  & -251.34   & -253.69 & 7.24  & 7.47 &-243.71 & -230    \\            
                      &         & (-251.38) & (-251.57)& (-251.55) &         &       &      &        & (-243)  \\            
\ce{E{-}CH3CHNH + H}  & -213.82 & -222.30   & -221.99  & -221.92   & -224.10 & 4.00  & 4.14 &-218.16 & -202    \\            
                      &         & (-222.65) & (-222.34)& (-222.27) &         &       &      &        & (-217)  \\            
\ce{Z{-}CH3CHNH + H}  & -210.94 & -219.52   & -219.22  & -219.13   & -221.31 & 4.06  & 4.20 &-215.32 & -199    \\            
                      &         & (-219.85) & (-219.55)& (-219.46) &         &       &      &        & (-214)  \\            
\ce{C2H4 + NH2}       & -242.67 & -248.70   & -        & -         & -250.94 & 8.26  & 8.64 &-240.06 & -228    \\            
                      &         & (-248.78) &          &           &         &       &      &        &  (-241) \\ \bottomrule
\end{tabular}
\end{adjustbox}
\label{tab:energie}

\textsuperscript{\emph{a}} CBS+CV scheme: the $n$=T,Q set of bases for the extrapolation to the CBS limit and cc-pCVTZ for the the CV contribution. Within parentheses, the results for the $n$=Q,5 set (CBS) and cc-pCVQZ (CV). \\
\textsuperscript{\emph{b}} CBS+CV: $n$=T,Q set (CBS) and cc-pCVTZ (CV). Within parentheses, $n$=Q,5 (CBS) and cc-pCVQZ (CV). \\
\textsuperscript{\emph{c}} Relative ZPVE corrections at the B2 level. \\
\textsuperscript{\emph{d}} CBS+CV electronic energies augmented by anharmonic ZPVE corrections. \\
\textsuperscript{\emph{e}} Data from ref. \cite{balucani2018}: values at the CCSD(T)/aug-cc-pVTZ level, within parentheses W1 values. ZPVE corrected values. \\
\end{table}

From the inspection of Table \ref{tab:energie}, it is evident that the CBS+CV energies obtained by extrapolating the correlation energy using either cc-pVTZ/cc-pVQZ or cc-pVQZ/cc-pV5Z are in close agreement, the largest difference being around \SI{0.4}{\kilo\joule\per\mol} for TS7. Incorporation of the fT and pQ contributions provides negligible effects, the only exceptions being TS2, TS3 and TS4, for which deviations of about 2-\SI{3}{\kilo\joule\per\mol} are noted. As expected from our previous experience \cite{D0CP00561D,Salta2000,Lupi2020}, the results obtained by using the ChS model are accurate. In fact, they are within \SI{2.5}{\kilo\joule\per\mol} from the corresponding CBS+CV values, and within \SI{2}{\kilo\joule\per\mol} with respect to CBS+CV+fT+pQ results. Finally, by incorporating the DBOC and relativistic corrections, the HEAT-like values have been obtained. These negligibly differ from the corresponding CBS+CV+fT+pQ results, the sum of the two contributions being on the order of 0.1-\SI{0.2}{\kilo\joule\per\mol}.

In Table \ref{tab:energie}, the $\Delta H_0^0$ values correspond to the CBS+CV energies augmented by anharmonic B2 ZPVEs. It is noted that they are in agreement with those calculated in ref. \cite{balucani2018} at the W1 level, even if available for a lesser number of species. On the other hand, the CCSD(T)/aug-cc-pVTZ level, 
employed in ref. \cite{balucani2018} for all stationary points, shows deviations as large as \SI{19}{\kilo\joule\per\mol} with respect to our CBS+CV results, with an average overestimation of about \SI{14.5}{\kilo\joule\per\mol}. It is particular interesting that the B2 model chemistry performs better than CCSD(T)/aug-cc-pVTZ computations, indeed the largest deviation of the former from CBS+CV data is \SI{9}{\kilo\joule\per\mol}, systematically overestimating the latter results.

\subsection{The \texorpdfstring{\ce{NH + C2H5}}{NH + C2H5} reaction: Rate Coefficients}

\begin{figure}[H]
\centering
\includegraphics[width=0.85\textwidth]{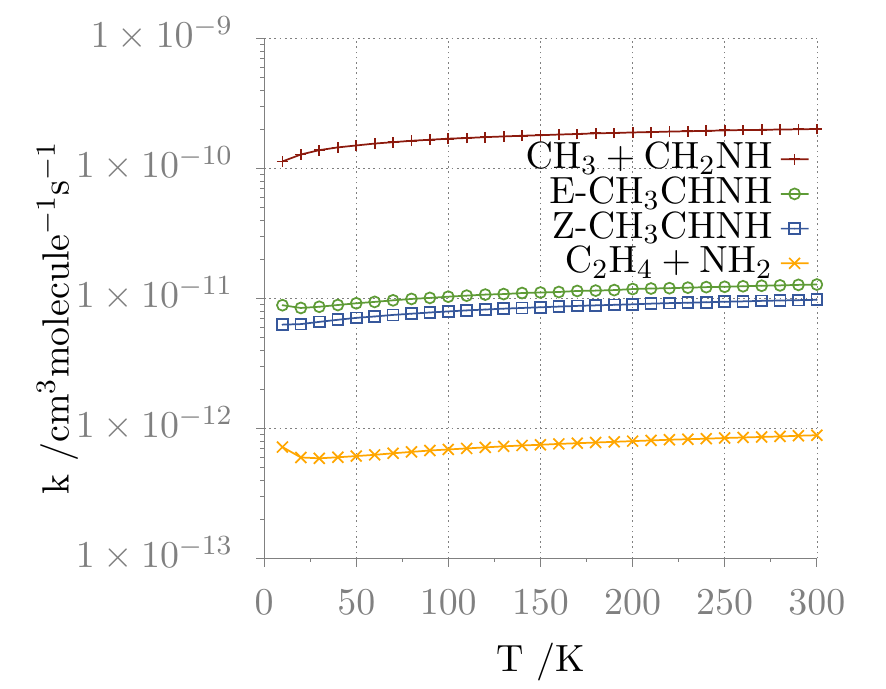}
\caption{Rate coefficients as a function of temperature for the four products.}\label{fig:rate}
\end{figure}

As described in Section \ref{sec:kinmod}, for the reaction paths shown in Figure \ref{fig:pes}, channel specific rate constants have been computed using AITSTME. For these calculations, the CBS+CV energies, corrected by anharmonic ZPVEs, have employed for all intermediate and transition states of the PES. The rate coefficients have been evaluated in the 10-300 K temperature range and at pressure of \num{1d-12} atm, the results being collected in Table \ref{tab:rate}. The corresponding temperature dependence plots are shown in Figure \ref{fig:rate}. 

The kinetic analysis shows that, in agreement with previous computations, only the channels leading to methanimine (R1), E-ethanimine (R2a), and Z-ethanimine (R2b) are significant. The contribution of the channel leading to \ce{C2H4 + NH2} is indeed negligible because of the limitations explained above. Focusing on the temperature dependence of the rate constant, an interesting feature can be noted at very low temperature (i.e. 10 K). In fact, at such low temperatures, the trend of $k$ for the channel leading to \ce{C2H4 + NH2} shows a larger variation with temperature than the other paths. This can be explained as a quantum tunneling effect because the TSs of this path are higher in energy and the rearrangements involve H transfer. 

As mentioned in Section \ref{r-pes}, the two reactants, \ce{CH3CH2} and NH, are in a spin-doublet and a spin-triplet state, respectively. As a consequence, this results in a total of six spin-microstates, which can be classified in reactive (overall: spin-doublet state) and non-reactive (overall: spin-quartet state). Since the number of the reactive spin-microstates is 2 and that of the non-reactive ones is 4, a statistical factor of 1/3 must be applied to the total reaction rate constant in order to account for this ratio. This has been indeed taken into account in our calculations.

\begin{table}[ht]
\centering
\caption{Product-formation rate constants (in cm$^3$ molecule$^{-1}$ s$^{-1}$) at \num{1d-12} atm as a function of the temperature.}\label{tab:rate}
\begin{tabular}{@{}lllll@{}}
\toprule
T(K) & \ce{CH3 + CH2NH} & \ce{E{-}CH3CHNH} & \ce{Z{-}CH3CHNH} & \ce{C2H4 + NH2} \\ \midrule
10   & \num{1.13d-10}  & \num{8.87d-12}  & \num{6.29d-12} & \num{7.19d-13}\\
20   & \num{1.28d-10}  & \num{8.45d-12}  & \num{6.35d-12} & \num{5.98d-13}\\
30   & \num{1.38d-10}  & \num{8.64d-12}  & \num{6.61d-12} & \num{5.90d-13}\\
40   & \num{1.45d-10}  & \num{8.91d-12}  & \num{6.85d-12} & \num{6.00d-13}\\
50   & \num{1.50d-10}  & \num{9.18d-12}  & \num{7.08d-12} & \num{6.14d-13}\\
60   & \num{1.55d-10}  & \num{9.41d-12}  & \num{7.27d-12} & \num{6.27d-13}\\
70   & \num{1.59d-10}  & \num{9.67d-12}  & \num{7.47d-12} & \num{6.45d-13}\\
80   & \num{1.63d-10}  & \num{9.90d-12}  & \num{7.64d-12} & \num{6.61d-13}\\
90   & \num{1.66d-10}  & \num{1.01d-11}  & \num{7.80d-12} & \num{6.76d-13}\\
100  & \num{1.69d-10}  & \num{1.03d-11}  & \num{7.95d-12} & \num{6.90d-13}\\
110  & \num{1.71d-10}  & \num{1.05d-11}  & \num{8.08d-12} & \num{7.03d-13}\\
120  & \num{1.74d-10}  & \num{1.07d-11}  & \num{8.21d-12} & \num{7.16d-13}\\
130  & \num{1.76d-10}  & \num{1.08d-11}  & \num{8.33d-12} & \num{7.28d-13}\\
140  & \num{1.78d-10}  & \num{1.10d-11}  & \num{8.44d-12} & \num{7.39d-13}\\
150  & \num{1.80d-10}  & \num{1.11d-11}  & \num{8.55d-12} & \num{7.50d-13}\\
160  & \num{1.82d-10}  & \num{1.12d-11}  & \num{8.65d-12} & \num{7.60d-13}\\
170  & \num{1.84d-10}  & \num{1.14d-11}  & \num{8.75d-12} & \num{7.70d-13}\\
180  & \num{1.86d-10}  & \num{1.15d-11}  & \num{8.84d-12} & \num{7.80d-13}\\
190  & \num{1.87d-10}  & \num{1.16d-11}  & \num{8.93d-12} & \num{7.90d-13}\\
200  & \num{1.89d-10}  & \num{1.18d-11}  & \num{9.02d-12} & \num{7.99d-13}\\
210  & \num{1.90d-10}  & \num{1.19d-11}  & \num{9.11d-12} & \num{8.09d-13}\\
220  & \num{1.92d-10}  & \num{1.20d-11}  & \num{9.19d-12} & \num{8.18d-13}\\
230  & \num{1.93d-10}  & \num{1.21d-11}  & \num{9.27d-12} & \num{8.26d-13}\\
240  & \num{1.94d-10}  & \num{1.22d-11}  & \num{9.35d-12} & \num{8.35d-13}\\
250  & \num{1.96d-10}  & \num{1.23d-11}  & \num{9.42d-12} & \num{8.44d-13}\\
260  & \num{1.97d-10}  & \num{1.24d-11}  & \num{9.49d-12} & \num{8.52d-13}\\
270  & \num{1.98d-10}  & \num{1.25d-11}  & \num{9.57d-12} & \num{8.61d-13}\\
280  & \num{1.99d-10}  & \num{1.26d-11}  & \num{9.64d-12} & \num{8.69d-13}\\
290  & \num{2.00d-10}  & \num{1.27d-11}  & \num{9.71d-12} & \num{8.78d-13}\\
300  & \num{2.01d-10}  & \num{1.28d-11}  & \num{9.77d-12} & \num{8.86d-13}\\ \bottomrule
\end{tabular}
\end{table}

\begin{table}[h!]
\centering
\caption{Product branching ratios at various temperatures.}
\label{tab:br}
\begin{tabular}{@{}ccccc@{}}
\toprule
Branching ratios & \ce{CH3 + CH2NH} & \ce{E{-}CH3CHNH + H} & \ce{Z{-}CH3CHNH + H} & \ce{C2H4 + NH2} \\ \midrule
10 K    & 87.7 \%	  & 6.9	\%       & 4.9 \%	    &  0.6 \% \\
100 K	& 89.9 \%	  &5.5  \%	     &4.2  \%       &  0.4 \%  \\
300	K   & 89.6 \%     &5.7	\%       &4.3  \%	    &  0.4 \%   \\    \bottomrule
\end{tabular}
\end{table}

The branching ratios at selected temperatures are shown in Table \ref{tab:br}. Is clear that, at all the temperatures considered, the channel leading to \ce{CH3 + CH2NH} is by far the dominant one, indeed accounting for more than 87\% of the total yield. The ratio between the E- and Z- isomers of ethanimine spans from 1.3 to 1.4 in the entire analyzed range. Even if this value is greater than that obtained by Balucani \emph{et al}. \cite{balucani2018}, it is still underestimated with respect to the factor of ca. 3 derived from the observation of Loomis \emph{et al.} \cite{loomis2013}. While our theoretical estimate is difficult to improve (because already at the state of the art), this discrepancy deserves to be addressed in some details. First of all, the astronomical value is affected by a large uncertainty mostly due to the fact that, for the analysis of E-ethanimine features, the authors had to fix its excitation temperature to that derived for Z-\ce{CH3CHNH}, this resulting in an approximate value of the column density of the E species (see ref. \cite{loomis2013} for a thorough account). Second, a non negligible contribution of dust-grain chemistry to the production of ethanimine cannot be excluded.

Finally, kinetic calculations have also been performed using the CBS+CV+fT+pQ energies, corrected by anharmonic B2 ZPVEs. In these calculations, the reaction channel R3 has been excluded, since it has been demonstrated that it gives non relevant contribution to the total reactive flux. The results are almost identical to those obtained exploiting CBS+CV energies. For the sake of completeness, they are only reported in the SI.

\subsection{\texorpdfstring{E-/Z-\ce{CH3CHNH}}{E-/Z-CH3CHNH}: Spectroscopic characterization}

In ref. \cite{melli2018}, an integrated experimental-theoretical strategy for the spectroscopic characterization required for guiding astronomical searches of molecular signatures in space has been presented together with its application to ethanimine. To exploit the interplay of experiment and theory, we have relied on the VMS-ROT software \cite{Lic17}. 

Despite the fact that state-of-the-art QC computations are able to provide predictions for rotational transitions with an accuracy better than 0.1\% (see, e.g., ref. \cite{rotstat} and Table \ref{tab-spec}), this is not generally sufficient for guiding astronomical searches and/or assignments, thus requiring experimental determinations of the corresponding spectroscopic parameters. In fact, when dealing with broadband unbiased astronomical surveys, rotational transition frequencies need to be known with an accuracy preferably better than 100 kHz, an accuracy that can be easily obtained from experimental studies.

\begin{table}[t!]
\centering
\caption{Computed and experimental rotational parameters (values in MHz) of ethanimine.}\label{tab-spec}
\begin{tabular}{lccccccc} \hline 
& \multicolumn{3}{c}{E-\ce{CH3CHNH}} && \multicolumn{3}{c}{Z-\ce{CH3CHNH}}\\
             & \multicolumn{2}{c}{Theory}     & Experiment$^a$                 && \multicolumn{2}{c}{Theory}     & Experiment$^a$ \\ \cline{2-3} \cline{6-7}
             &   Best estimates$^b$  &   B2$^c$          &                         && Best estimates$^b$    &   B2$^c$         &                             \\ \hline                                
$A_0$        &53178.26               & 53398.97          &   53120.561(30)         &&50002.63               & 50288.21          &49964.87(93)                 \\                      
$B_0$        & 9780.14               & 9744.02           &    9782.7720(47)        && 9831.98               & 9781.22           & 9832.4823(96)               \\                       
$C_0$        & 8702.82               & 8679.22           &    8697.0263(46)        && 8652.81               & 8621.21           & 8646.0305(94)               \\                 
$\Delta_{J}$ &    6.48$\times10^{-3}$&6.39$\times10^{-3}$&6.4641(49)$\times10^{-3}$&&    6.99$\times10^{-3}$&6.96$\times10^{-3}$&    6.938(13)$\times10^{-3}$ \\       
$\Delta_{K}$ &    0.568              & 0.575             &       0.5763(34)        &&    0.468              & 0.480             &    0.468$^d$                \\                           
$\Delta_{JK}$&   -0.0165             & -0.0165           &      -0.01403(21)       &&   -0.0163             & -0.0182           &  -0.01219(23)               \\                          
$\delta_{J}$ &    1.09$\times10^{-3}$&1.06$\times10^{-3}$&1.1033(19)$\times10^{-3}$&&    1.25$\times10^{-3}$&1.24$\times10^{-3}$&    1.2657(65)$\times10^{-3}$\\
$\delta_{K}$ &   -0.0535             & -0.0518           &      -0.06709(59)       &&   -0.0522             & -0.0464           &   -0.0642(19)               \\                       
\\                                                                                                                                      $V_{3}\;^e$  & 563.1                 &  488.8            & 566.37(20)              && 523.3                 &  446.4            & 517.41(33)                  \\ \hline                                                 
  \end{tabular}

 \scriptsize{
\textsuperscript{\emph{a}} Watson A-reduction. Values in parenthesis denote one standard deviation and apply to the last digits of the constants. \\
\textsuperscript{\emph{b}} Equilibrium CBS+CV rotational constants augmented by B2 vibrational corrections. Quartic centrifugal distortion constants at the CCSD(T)/cc-pCVQZ level. For details, see \cite{melli2018}. \\
\textsuperscript{\emph{c}} This work. 
\textsuperscript{\emph{d}} Fixed at the best-estimated theoretical value. 
\textsuperscript{\emph{e}} $V_{3}$ values in cm\textsuperscript{-1}.
  }
\end{table}

\begin{figure}[H]
\centering 
\includegraphics[width=0.95\textwidth]{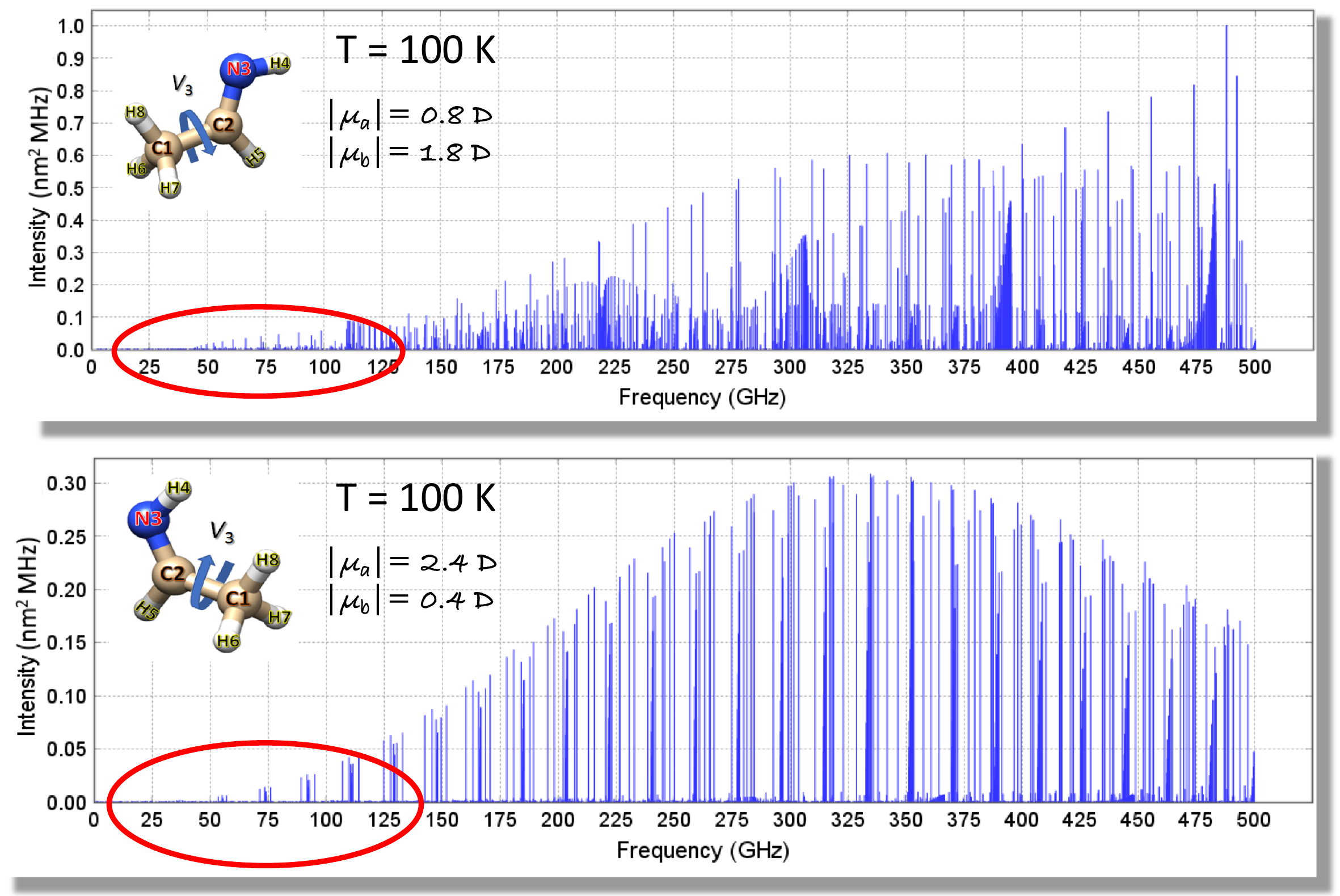}
\caption{Simulation of the rotational spectra of E- (top panel) and Z-\ce{CH3CHNH} (bottom panel) at T = 100 K obtained with VMS-ROT. The absolute values of the dipole moment components are also reported. The red circles highlight the frequency ranges for which measurements were available prior to the investigation of ref. \cite{melli2018}.}
\label{fig-spec}
\end{figure}

The spectroscopic parameters of the two isomers of ethanimine have been computed as explained in Section \ref{sec:spec-det} and, in more details, in ref. \cite{melli2018}. Table \ref{tab-spec} collects a selection of computed spectroscopic parameters, which are compared with their experimental counterparts (see ref. \cite{melli2018} for an extended version of this table). 
In order to improve the prediction of the rotational spectrum, the computed rotational constants have been replaced by those available from ref. \cite{lovas1980}. 
The resulting simulated spectra, shown in Figure \ref{fig-spec}, have then used to guide rotational spectrum measurements, thus leading to a straightforward assignment of the rotational transitions up to 300 GHz \cite{melli2018}.  

Back to Table \ref{tab-spec}, there is excellent agreement between the computed and experimental rotational constants, with the maximum discrepancy (i.e. 0.1\%) noted for the $A_0$ constant. For quartic centrifugal distortion constants, a quantitative agreement between theory and experiment is apparent as well. In Table \ref{tab-spec}, the determination of the $V_3$ term is also reported. As briefly mentioned in Section \ref{sec:spec-det}, this is due to the internal rotation of the \ce{-CH3} group (see Figures \ref{fig:v3} and \ref{fig-spec}), which leads to a periodic potential energy along the torsional angle, with $V_3$ being the energy barrier (the reader is referred to ref. \cite{melli2018} for a detailed account). This parameter is important in the prediction and analysis of the rotational spectrum because, the $V_3$ barrier being finite, a tunneling effect occurs. This leads to a splitting of the threefold degeneracy into two levels, a nondegenerate A level and a doubly degenerate E level. As a consequence, rotational transitions in the A and E torsional sublevels are observed, and their separation depends on the value of $V_3$.

In Table \ref{tab-spec}, for comparison purposes, the results at the B2 level are also reported. Despite the reduced computational cost, the B2 parameters are accurate: the discrepancies for the rotational constants range, in relative terms, from 0.2\% to 0.5\%. The worsening in the prediction of the centrifugal distortion constants (with respect to the all-CCSD(T)/cc-pCVQZ level) is nearly negligible and, for the $V_3$ barrier, the disagreement is well within \SI{1}{\kilo\joule\per\mol}. These results are particularly significant for the study of larger systems, which are currently not amenable to the most accurate computations reported in the present study.

\section{Conclusions}

Astrochemistry is an interdisciplinary field involving chemistry, physics, and astronomy, and is strongly interconnected to astrobiology. The main aim of astrochemistry is to understand the chemical evolution of the universe, thus investigating the formation and destruction of molecules in space, but also their interaction with radiation and their feedback on physics of the environments. In the present contribution, the paradigmatic case of ethanimine has been used to drive the readers in a fascinating journey among the different computational tools presently available to investigate two major challenges in astrochemistry: the discovery of new molecules in space and the understanding of how they can be formed. Ethanimine has been chosen because of its prebiotic potential, thus being a first step toward the demonstration of the presence of biological building-block molecules in the ISM. 

Because of difficulties in experimentally mimicking the extreme conditions that characterize the interstellar medium, accurate state-of-the-art computational approaches play a fundamental role. We have demonstrated how state-of-the-art QC methodologies are able to accurately describe reactive PESs, to be complemented by kinetic calculations, and predict the spectroscopic parameters of the reaction products. 
Even if our state-of-the-art investigation on the gas-phase formation of ethanimine is not able to perfectly reproduce the [\ce{E{-}CH3CHNH}]/[\ce{Z{-}CH3CHNH}] ratio of ca. 3 issuing from astronomical observations, the result obtained (1.4) is of the correct order of magnitude and it can be considered as a satisfactory value in view of the complexity of the chemistry in space, involving several reactions at the same time both in the gas phase and on dust-grains. From a spectroscopic point of view, it has been shown how state-of-the-art QC methodologies can guide, support and complement rotational spectroscopy studies, the latter providing the rest frequencies required by astronomical searches.


\vspace{6pt} 



\funding{This research was funded by the Italian MIUR grant numbers 2017A4XRCA (PRIN 2017) and 2015F59J3R (PRIN 2015), by the University of Bologna (RFO funds), and by Scuola Normale Superiore (Grant number SNS18\_B\_TASINATO).}

\acknowledgments{The SMART@SNS Laboratory (http://smart.sns.it) is acknowledged for providing high-performance computer facilities}

\conflictsofinterest{The authors declare no conflict of interest.} 

\appendixtitles{no} 



\externalbibliography{yes}
\bibliography{Bibliografia}
\end{document}


\section{Geometrical parameters}
\begin{scriptsize}

\begin{figure}[H]
\centering
\includegraphics[width=0.8\textwidth]{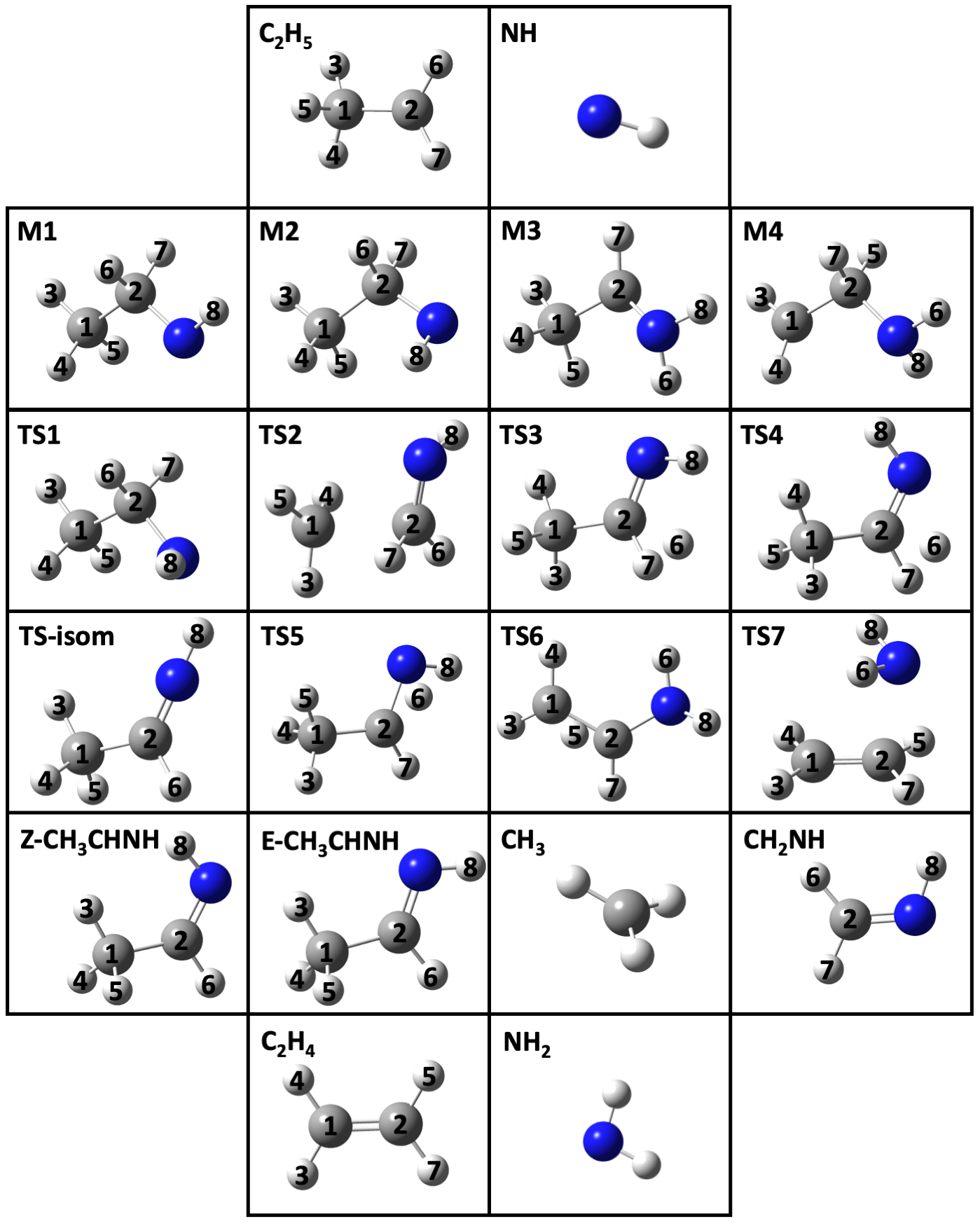}
\caption{Structures of the species appearing in the formation pathway of ethanimine as obtained at B2PLYP-D3(BJ)/aug-cc-pVTZ level of theory. Numeric labels are used for structural analysis in Table \ref{tab:geom}}\label{fig:specieSI}
\end{figure}

\clearpage

\begin{longtable}{@{}ccccc@{}}
\caption{Structural parameters of the stationary points of the \ce{NH + C2H5} reaction at different levels of theory. Distances are in \SI{}{\angstrom}, angles in \SI{}{\degree}.}\label{tab:geom} \\
\toprule
                          &             & B3LYP-D3    & \multicolumn{2}{c}{B2PLYP-D3(BJ)} \\* \midrule
\endfirsthead
\multicolumn{5}{c}%
{{\bfseries Table \thetable\ continued from previous page}} \\
\endhead
\cmidrule(l){2-5}
\endfoot
\endlastfoot
                          &             & aug-cc-pVTZ & aug-cc-pVTZ   & aug-cc-pVQZ   \\\midrule
\multirow{1}{*}{NH}       & r(NH)       & 1.040       & 1.036         & 1.035 \\\midrule
\multirow{7}{*}{\ce{C2H5}} & r(C2H6) & 1.080          & 1.078          & 1.078\\
                           & r(C2H7) & 1.080          & 1.080          & 1.080\\
                           & r(C1C2) & 1.484 & 1.484 & 1.483\\
                           & $\varphi$(H6C2C1) & 120.9 & 120.9 & 120.9\\
                           & $\varphi$(H3C1C2) & 111.9 & 111.7 & 111.7\\
                           & $\varphi$(H3C1H5) & 106.3 & 106.5 & 106.5\\
                           & $\varphi$(H5C1H4) & 106.3 & 106.5 & 106.5\\
                           & $\varphi$(H6C2H7) & 117.7 & 117.7 & 117.7\\
                           & $\vartheta$(H3C1C2H6) & 33.0 & 33.5 & 33.4\\
                           & $\vartheta$(H5C1C2H7  & 86.1 & 85.7 & 85.8\\\midrule
\multirow{12}{*}{M1}      & r(NH8)      & 1.025       & 1.023         & 1.022         \\
                          & r(C2N)      & 1.439       & 1.440         & 1.439         \\
                          & r(C1C2)     & 1.522       & 1.519         & 1.518         \\
                          & r(C1H3)     & 1.091       & 1.089         & 1.089         \\
                          & r(C1H4)     & 1.090       & 1.089         & 1.088         \\
                          & r(C1H5)     & 1.090       & 1.089         & 1.088         \\
                          & r(C2H6)     & 1.102       & 1.099         & 1.099         \\
                          & r(C2H7)     & 1.102       & 1.099         & 1.099         \\
                          & $\varphi$(C2NH8)   & 107.6       & 107.3         & 107.3         \\
                          & $\vartheta$(C1C2NH8) & 180.0       & -180.0        & -180.0        \\
                          & $\vartheta$(H7C2NH8) & 57.0        & 57.2          & 57.2          \\
                          & $\vartheta$(H6C2NH8) & -57.0       & -57.2         & -57.2         \\\midrule
\multirow{12}{*}{M2}      & r(NH8)      & 1.026       & 1.024         & 1.023         \\
                          & r(C2N)      & 1.440       & 1.441         & 1.440         \\
                          & r(C1C2)     & 1.534       & 1.530         & 1.529         \\
                          & r(C1H3)     & 1.091       & 1.089         & 1.088         \\
                          & r(C1H4)     & 1.092       & 1.090         & 1.089         \\
                          & r(C1H5)     & 1.090       & 1.089         & 1.088         \\
                          & r(C2H6)     & 1.103       & 1.100         & 1.099         \\
                          & r(C2H7)     & 1.092       & 1.090         & 1.089         \\
                          & $\varphi$(C2NH8)   & 107.3       & 106.8         & 106.9         \\
                          & $\vartheta$(C1C2NH8) & -46.3       & -49.0         & -49.1         \\
                          & $\vartheta$(H7C2NH8) & -170.4      & -172.6        & -172.7        \\
                          & $\vartheta$(H6C2NH8) & 74.9        & 71.9          & 71.8          \\\midrule
\multirow{14}{*}{M3}      & r(NH8)      & 1.008            & 1.008         &    1.007           \\
                          & r(NH6)      & 1.010            & 1.009         &    1.008           \\
                          & r(C2N)      & 1.398            & 1.398         &    1.396           \\
                          & r(C1C2)     & 1.487            & 1.487       &      1.486         \\
                          & r(C1H3)     & 1.089            & 1.088       &      1.087         \\
                          & r(C1H4)     & 1.101            & 1.098       &      1.098         \\
                          & r(C1H5)     & 1.096            & 1.093       &      1.093         \\
                          & r(C2H7)     & 1.082            & 1.080       &      1.080         \\
                          & $\varphi$(C2NH6)   & 115.2            & 114.7         &  114.8              \\
                          & $\varphi$(C2NH8)   & 115.5            & 115.1         &  115.3             \\
                          & $\vartheta$(C1C2NH6) & 40.3            & 41.7          &   41.4            \\
                          & $\vartheta$(C1C2NH8) & 172.1           & 172.3         &  172.4               \\
                          & $\vartheta$(H7C2NH6) & -170            & -170.2        &  -170.3             \\
                          & $\vartheta$(H7C2NH8) & -38.1            & -39.5         &  -39.2             \\\midrule
\multirow{16}{*}{M4}      & r(NH8)      &  1.013           & 1.012         &   1.011            \\
                          & r(NH6)      &  1.012           & 1.011         &   1.010            \\
                          & r(C2N)      &  1.467           & 1.466         &   1.464            \\
                          & r(C1C2)     &  1.482           & 1.482         &   1.481            \\
                          & r(C1H3)     &  1.080           & 1.078         &   1.078            \\
                          & r(C1H4)     &  1.081           & 1.079         &   1.078            \\
                          & r(C2H7)     &  1.101           & 1.098         &   1.098            \\
                          & r(C2H5)     &  1.101                & 1.098         &  1.097             \\
                          & $\varphi$(C2NH6)   &  111.1           & 110.8         &  110.9             \\
                          & $\varphi$(C2NH8)   &  110.6           & 110.3         &   110.3            \\
                          & $\vartheta$(C1C2NH6) &  -175.8           & -176.4        &  -176.2             \\
                          & $\vartheta$(C1C2NH8) &  64.8           & 65.0          &    65.0           \\
                          & $\vartheta$(H5C2NH6) &  59.8           & 59.6          &   59.8            \\
                          & $\vartheta$(H7C2NH6) &  -55.9           & -56.6         &  -56.4             \\
                          & $\vartheta$(H5C2NH8) &  -59.5           & -59.0         &  -59.0             \\
                          & $\vartheta$(H7C2NH8) &  -175.3           & -175.2        & -175.2              \\\midrule
\multirow{12}{*}{TS1}     & r(NH8)      & 1.023       & 1.021         & 1.021         \\
                          & r(C2N)      & 1.445       & 1.446         & 1.446         \\
                          & r(C1C2)     & 1.539       & 1.534         & 1.534         \\
                          & r(C1H3)     & 1.091       & 1.089         & 1.089         \\
                          & r(C1H4)     & 1.090       & 1.089         & 1.089         \\
                          & r(C1H5)     & 1.089       & 1.088         & 1.088         \\
                          & r(C2H6)     & 1.095       & 1.093         & 1.093         \\
                          & r(C2H7)     & 1.097       & 1.094         & 1.094         \\
                          & $\varphi$(C2NH8)   & 109.0       & 108.9         & 108.9         \\
                          & $\vartheta$(C1C2NH8) & -114.2      & -114.6        & -114.6        \\
                          & $\vartheta$(H7C2NH8) & 129.0       & 128.6         & 128.6         \\
                          & $\vartheta$(H6C2NH8) & 9.7         & 9.1           & 9.1           \\\midrule
\multirow{12}{*}{TS2}     & r(NH8)      & 1.020       & 1.019         & 1.019         \\
                          & r(C2N)      & 1.293       & 1.293         & 1.293         \\
                          & r(C1C2)     & 2.271       & 2.224         & 2.224         \\
                          & r(C1H3)     & 1.081       & 1.080         & 1.080         \\
                          & r(C1H4)     & 1.080       & 1.078         & 1.078         \\
                          & r(C1H5)     & 1.079       & 1.078         & 1.078         \\
                          & r(C2H6)     & 1.091       & 1.090         & 1.090         \\
                          & r(C2H7)     & 1.087       & 1.085         & 1.085         \\
                          & $\varphi$(C2NH8)   & 110.8       & 110.5         & 110.5         \\
                          & $\vartheta$(C1C2NH8) & -89.8       & -89.5         & -89.5         \\
                          & $\vartheta$(H7C2NH8) & 172.7       & 172.7         & 172.7         \\
                          & $\vartheta$(H6C2NH8) & 10.8        & 11.0          & 11.0          \\\midrule
\multirow{12}{*}{TS3}     & r(NH8)      & 1.019       & 1.018         & 1.018         \\
                          & r(C2N)      & 1.284       & 1.286         & 1.286         \\
                          & r(C1C2)     & 1.501       & 1.500         & 1.500         \\
                          & r(C1H3)     & 1.090       & 1.088         & 1.088         \\
                          & r(C1H4)     & 1.088       & 1.086         & 1.086         \\
                          & r(C1H5)     & 1.094       & 1.092         & 1.092         \\
                          & r(C2H6)     & 1.926       & 1.829         & 1.829         \\
                          & r(C2H7)     & 1.096       & 1.094         & 1.094         \\
                          & $\varphi$(C2NH8)   & 111.1       & 110.7         & 110.7         \\
                          & $\vartheta$(C1C2NH8) & 175.0       & 174.2         & 174.2         \\
                          & $\vartheta$(H7C2NH8) & 7.6         & 8.3           & 8.3           \\
                          & $\vartheta$(H6C2NH8) & -79.5       & -80.8         & -80.8         \\\midrule
\multirow{12}{*}{TS4}     & r(NH8)      & 1.022       & 1.021         & 1.021         \\
                          & r(C2N)      & 1.283       & 1.285         & 1.285         \\
                          & r(C1C2)     & 1.508       & 1.506         & 1.506         \\
                          & r(C1H3)     & 1.090       & 1.088         & 1.088         \\
                          & r(C1H4)     & 1.093       & 1.092         & 1.092         \\
                          & r(C1H5)     & 1.090       & 1.089         & 1.089         \\
                          & r(C2H6)     & 1.091       & 1.090         & 1.090         \\
                          & r(C2H7)     & 1.927       & 1.830         & 1.830         \\
                          & $\varphi$(C2NH8)   & 110.9       & 110.4         & 110.4         \\
                          & $\vartheta$(C1C2NH8) & 8.7         & 9.6           & 9.6           \\
                          & $\vartheta$(H7C2NH8) & -97.2       & -96.4         & -96.4         \\
                          & $\vartheta$(H6C2NH8) & 175.7       & 175.2         & 175.2         \\\midrule
\multirow{10}{*}{TS-isom} & r(NH8)      & 0.987       & 0.987         & 0.987         \\
                          & r(C2N)      & 1.234       & 1.237         & 1.237         \\
                          & r(C1C2)     & 1.512       & 1.510         & 1.510         \\
                          & r(C1H3)     & 1.088       & 1.087         & 1.087         \\
                          & r(C1H4)     & 1.094       & 1.092         & 1.092         \\
                          & r(C1H5)     & 1.094       & 1.092         & 1.092         \\
                          & r(C2H6)     & 1.110       & 1.107         & 1.107         \\
                          & $\varphi$(C2NH8)   & 179.4       & 179.4         & 179.4         \\
                          & $\vartheta$(C1C2NH8) & -0.1        & 0.6           & 0.6           \\
                          & $\vartheta$(H6C2NH8) & 179.9       & -179.4        & -179.4        \\\midrule
\multirow{13}{*}{TS5}     & r(NH8)      &  1.020           & 1.020         &   1.019       \\
                          & r(NH6)      &  1.252           & 1.251         &   1.251       \\
                          & r(C2N)      &  1.452           & 1.453         &   1.451       \\
                          & r(C1C2)     &  1.494           & 1.492         &   1.491       \\
                          & r(C1H3)     &  1.090           & 1.089         &   1.088       \\
                          & r(C1H4)     &  1.097           & 1.095         &   1.094          \\
                          & r(C1H5)     &  1.091           & 1.089         &   1.088       \\
                          & r(C2H6)     &  1.284           & 1.276         &   1.275       \\
                          & r(C2H7)     &  1.087           & 1.085         &   1.084       \\
                          & $\varphi$(C2NH8)   & 107.6   & 107.2         &   107.3       \\
                          & $\vartheta$(C1C2NH8) & -165.1  & -165.6        &   -165.6     \\
                          & $\vartheta$(H7C2NH8) & -14.0& -14.4            &   -14.4       \\
                          & $\vartheta$(H7C2NH6) & -106.2  & -106.3        &   -106.3      \\\midrule
\multirow{15}{*}{TS6}     & r(NH8)      &   1.009          & 1.009         &    1.008      \\
                          & r(NH6)      &   1.009          & 1.009         &    1.008      \\
                          & r(C2N)      &   1.413          & 1.413         &    1.411      \\
                          & r(C1C2)     &   1.483          & 1.481         &    1.480      \\
                          & r(C1H3)     &   1.080          & 1.078         &    1.077      \\
                          & r(C1H4)     &   1.080          & 1.079         &    1.078      \\
                          & r(C1H5)     &   1.326          & 1.319         &    1.318     \\
                          & r(C2H5)     &   1.306          & 1.300         &    1.299      \\
                          & r(C2H7)     &   1.083          & 1.081         &    1.080      \\
                          & $\varphi$(C2NH8)   &  114.7           & 114.4         &   114.5       \\
                          & $\varphi$(C2NH6)   &  113.9           & 113.3         &   113.4       \\
                          & $\vartheta$(C1C2NH8) &   145.5          & 144.1         &   144.1       \\
                          & $\vartheta$(C1C2NH6) &   16.8          & 16.6          &    16.3       \\
                          & $\vartheta$(H7C2NH8) &   -66.2          & -67.8         &   -67.6       \\
                          & $\vartheta$(H7C2NH6) &    165.1         & 164.8         &   164.6      \\\midrule
\multirow{14}{*}{TS7}     & r(NH8)      &  1.024           & 1.023         &   1.022       \\
                          & r(NH6)      &  1.024           & 1.023         &   1.022      \\
                          & r(C2N)      &  2.184           & 2.156         &   2.154       \\
                          & r(C1C2)     &  1.353           & 1.351         &   1.350      \\
                          & r(C1H3)     &  1.082           & 1.080         &   1.080         \\
                          & r(C1H4)     &  1.082           & 1.080         &   1.080     \\
                          & r(C2H5)     &  1.080           & 1.079         &   1.079      \\
                          & r(C2H7)     &  1.080           & 1.079         &   1-079      \\
                          & $\varphi$(C2NH8)   &   98.6          & 98.6          &  98.6        \\
                          & $\varphi$(C2NH6)   &   98.6          & 98.6          &  98.6         \\
                          & $\vartheta$(C1C2NH8) &   -52.5          & -52.4         &  -52.4        \\
                          & $\vartheta$(C1C2NH6) &    52.5       & 52.4          &      52.4    \\
                          & $\vartheta$(H7C2NH8) &    -174.6         & -174.5        &  -174.5       \\
                          & $\vartheta$(H7C2NH6) &  -69.7           & -69.6         &   -69.6      \\\midrule
\multirow{2}{*}{\ce{CH3}} & r(CH) & 1.078x3 & 1.076x3 & 1.075x3\\
                            & $\varphi$(HCH) & 120.0x3 & 120.0x3 &
                            120.0x3\\\midrule
\multirow{4}{*}{\ce{CH2NH}} & r(C2N) & 1.264 & 1.269 & 1.268\\
                            & $\varphi$(C2NH8) & 111.4 & 110.9 &
                            110.9\\
                            & $\varphi$(H6C2H7) & 116.1 & 116.7 & 116.7\\
                            & $\vartheta$(H6C2NH8) & 0.0 & 0.0 & 0.0\\
                            & $\vartheta$(H7C2NH8) & 180.0 & 180.0 & 180.0\\\midrule
\multirow{5}{*}{E-\ce{CH3CHNH}} & r(C2N) & 1.267 & 1.272 & 1.270\\
                                & r(C1C2) & 1.495 & 1.493 & 1.492\\
                                & r(C2H6) & 1.097 & 1.094 & 1.094\\
                                & $\varphi$(C2NH8) & 111.5 & 110.8 & 110.9\\
                                & $\vartheta$(C1C2NH8) & 180.0 & 180.0 & 180.0\\
                                & $\vartheta$(H6C2NH8) & 0.0 & 0.0 & 0.0\\\midrule
\multirow{5}{*}{Z-\ce{CH3CHNH}} & r(C2N) & 1.266 & 1.271 & 1.269\\
                                & r(C1C2) & 1.500 & 1.498 & 1.498\\
                                & r(C2H6) & 1.093 & 1.090 & 1.090\\
                                & $\varphi$(C2NH8) & 111.3 & 110.6 & 110.6\\
                                & $\vartheta$(C1C2NH8) & 0.0 & 0.0 & 0.0\\
                                & $\vartheta$(H6C2NH8) & 180.0 & 180.0  & 180.0\\\midrule
\multirow{2}{*}{\ce{NH2}}       & r(NH) & 1.027 & 1.024 & 1.023 \\
                                & $\varphi$(HNH) & 103.3 & 103.3 & 103.4 \\\midrule
\multirow{5}{*}{\ce{C2H4}}      & r(C1C2) & 1.325& 1.328 & 1.327 \\
                                & $\varphi$(HCC) & 121.7x3& 121.6x3 & 121.6x3\\
                                & $\varphi$(HCH) &116.5x2 & 116.9x2 &116.9x2 \\
                                & $\vartheta$(H4C1C2H5) & 0.0 & 0.0 & 0.0\\
                                & $\vartheta$(H4C1C2H7) & 180.0 & 180.0 & 180.0  \\* \bottomrule
\end{longtable}
\end{scriptsize}

\clearpage

\section{Absolute energies}
\begin{table}[ht!]
\caption{Energies are in \SI{}{\hartree}. Reference geometries at B2PLYP-D3(BJ)/aug-cc-pVTZ level of theory.}
\label{tab:absen}
\centering
\resizebox{\textwidth}{!}{%
\begin{tabular}{@{}ccccccc@{}}
\toprule
\multirow{2}{*}{} & \multirow{2}{*}{B2PLYP-D3(BJ)/aug-cc-pVTZ} & \multicolumn{2}{c}{CCSD(T)/CBS+CV} & \multicolumn{2}{c}{CCSD(T)/CBS+CV+fT+pQ} & \multirow{2}{*}{ChS} \\ \cmidrule(lr){3-6}
                  &                                            & T,Q              & Q,5             & T,Q                 & Q,5                &                      \\ \midrule
NH                & -55.2025220                                & -55.2143895      & -55.2190340     & -55.2149417         & -55.2195863        & -55.2141825          \\
\ce{C2H5}         & -79.1066887                                & -79.1401015      & -79.1476187     & -79.1411415         & -79.1486587        & -79.1405321          \\
M1                & -134.4406761                               & -134.4886002     & -134.5007278    & -134.4901229        & -134.5022506       & -134.4890983         \\
M2                & -134.4395633                               & -134.4875802     & -134.4997073    & -134.4891120        & -134.5012390       & -134.4881290         \\
M3                & -134.4550505                               & -134.5026174     & -134.5147721    & -                   & -                  & -134.5033022         \\
M4                & -134.4374715                               & -134.4862685     & -134.4983261    & -                   & -                  & -134.4870289         \\
TS1               & -134.4379875                               & -134.4859987     & -134.4981322    & -134.4875296        & -134.4996631       & -134.4865507         \\
TS2               & -134.3944707                               & -134.4406919     & -134.4529517    & -134.4433467        & -134.4556065       & -134.4413882         \\
TS3               & -134.3836749                               & -134.4314920     & -134.4437482    & -134.4337551        & -134.4460113       & -134.4319856         \\
TS4               & -134.3826904                               & -134.4305599     & -134.4428110    & -134.4328182        & -134.4450693       & -134.4310511         \\
TS-isom           & -133.8465558                               & -133.8919738     & -133.9042667    & -133.8934507        & -133.9057435       & -133.8929897         \\
TS5               & -134.3817405                               & -134.4296502     & -134.4417896    & -                   & -                  & -134.4305127         \\
TS6               & -134.3764047                               & -134.4239981     & -134.4361661    & -                   & -                  & -134.4249503         \\
TS7               & -134.3991008                               & -134.4450773     & -134.4574038    & -                   & -                  & -134.4456568         \\
\ce{CH2NH}        & -94.5902902                                & -94.6208211      & -94.6293610     & -94.6219034         & -94.6304432        & -94.6216334          \\
\ce{CH3}          & -39.8125589                                & -39.8293382      & -39.8330384     & -39.8299194         & -39.8336195        & -39.8297072          \\
\ce{E{-}CH3CHNH}  & -133.8919820                               & -133.9391388     & -133.9514338    & -133.9406117        & -133.9529067       & -133.9400468         \\
\ce{Z{-}CH3CHNH}  & -133.8908845                               & -133.9380799     & -133.9503673    & -133.9395581        & -133.9518454       & -133.9389855         \\
H                 & -0.4986682                                 & -0.5000222       & -0.5000222      & -0.5000222          & -0.5000222         & -0.5000222           \\
\ce{C2H4}         & -78.5427309                                & -78.5759224      & -78.5836340     & -                   & -                  & -78.5765986          \\
\ce{NH2}          & -55.8589088                                & -55.8732924      & -55.8777758     & -                   & -                  & -55.8736928          \\ \bottomrule
\end{tabular}%
}
\end{table}

\clearpage
\section{$\mathcal{T}_1$ diagnostic}

\begin{table}[ht!]
\caption{Coupled-cluster $\mathcal{T}_1$ diagnostic for all the species considered.}
\label{tab:my-table}
\centering
\begin{tabular}{@{}ccc@{}}
\toprule
          & \multicolumn{2}{c}{$\mathcal{T}_1$} \\ \cmidrule(l){2-3} 
          & cc-pVTZ    & cc-pVQZ   \\ \cmidrule(l){1-3} 
\ce{NH}        & 0.010      & 0.011     \\
\ce{C2H5}      & 0.010      & 0.010     \\
M1        & 0.012      & 0.013     \\
M2        & 0.012      & 0.013     \\
M3        & 0.014      & 0.014     \\
M4        & 0.010      & 0.011     \\
TS1       & 0.012      & 0.013     \\
TS2       & 0.031      & 0.031     \\
TS3       & 0.031      & 0.031     \\
TS4       & 0.031      & 0.031     \\
TS-isom   & 0.011      & 0.012     \\
TS5       & 0.016      & 0.016     \\
TS6       & 0.016      & 0.016     \\
TS7       & 0.031      & 0.031     \\
\ce{CH2NH}     & 0.012      & 0.012     \\
\ce{CH3}       & 0.008      & 0.009     \\
\ce{E{-}CH3CHNH} & 0.012      & 0.012     \\
\ce{Z{-}CH3CHNH} & 0.012      & 0.012     \\
\ce{C2H4}      & 0.011      & 0.011     \\
\ce{NH2}       & 0.008      & 0.009     \\ \bottomrule
\end{tabular}%

\end{table}

\section{CCSD(T)/CBS+CV+fT+pQ kinetic results}

\begin{figure}[H]
\centering
\includegraphics[width=\textwidth]{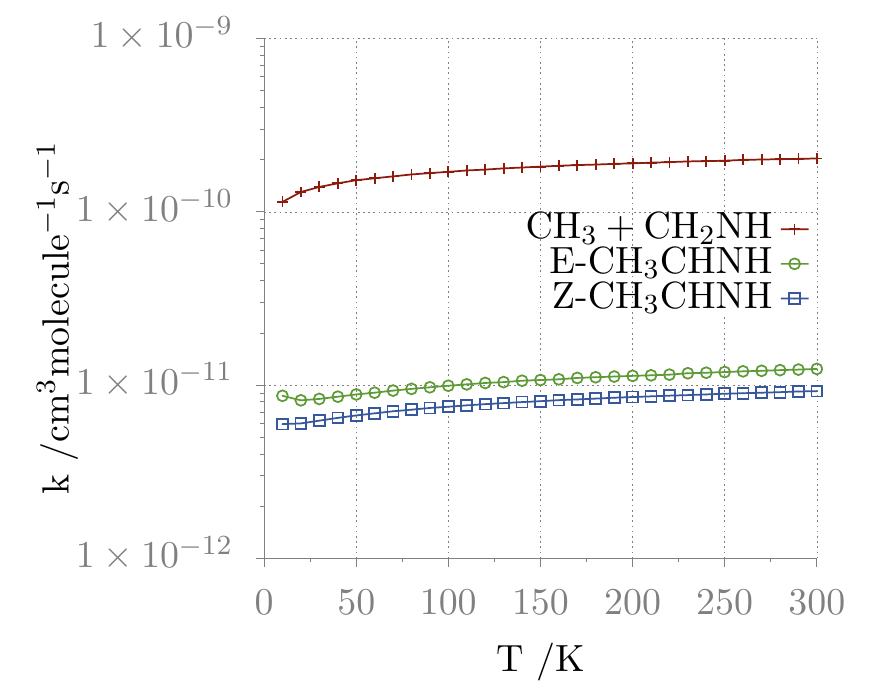}
\caption{Rate coefficients as a function of temperature.}\label{fig:rateftpq}
\end{figure}   

\begin{table}[ht]
\caption{Product-formation rate constants (in cm$^3$ molecule$^{-1}$ s$^{-1}$) at \num{1d-12} atm as a function of the temperature.}\label{tab:rate}
\centering
\begin{tabular}{@{}lllll@{}}
\toprule
T(K) & \ce{CH3 + CH2NH} & \ce{E{-}CH3CHNH} & \ce{Z{-}CH3CHNH} \\ \midrule
10   & \num{1.14d-10}  & \num{8.68d-12}  & \num{5.96d-12} \\
20   & \num{1.30d-10}  & \num{8.17d-12}  & \num{6.01d-12} \\
30   & \num{1.39d-10}  & \num{8.33d-12}  & \num{6.24d-12} \\
40   & \num{1.46d-10}  & \num{8.58d-12}  & \num{6.48d-12} \\
50   & \num{1.52d-10}  & \num{8.83d-12}  & \num{6.69d-12} \\
60   & \num{1.56d-10}  & \num{9.05d-12}  & \num{6.87d-12} \\
70   & \num{1.60d-10}  & \num{9.30d-12}  & \num{7.06d-12} \\
80   & \num{1.64d-10}  & \num{9.52d-12}  & \num{7.22d-12} \\
90   & \num{1.67d-10}  & \num{9.72d-12}  & \num{7.38d-12} \\
100  & \num{1.70d-10}  & \num{9.91d-12}  & \num{7.51d-12} \\
110  & \num{1.73d-10}  & \num{1.01d-11}  & \num{7.64d-12} \\
120  & \num{1.75d-10}  & \num{1.03d-11}  & \num{7.76d-12} \\
130  & \num{1.78d-10}  & \num{1.04d-11}  & \num{7.88d-12} \\
140  & \num{1.80d-10}  & \num{1.06d-11}  & \num{7.98d-12} \\
150  & \num{1.82d-10}  & \num{1.07d-11}  & \num{8.08d-12} \\
160  & \num{1.84d-10}  & \num{1.08d-11}  & \num{8.18d-12} \\
170  & \num{1.86d-10}  & \num{1.10d-11}  & \num{8.27d-12} \\
180  & \num{1.87d-10}  & \num{1.11d-11}  & \num{8.36d-12} \\
190  & \num{1.89d-10}  & \num{1.12d-11}  & \num{8.45d-12} \\
200  & \num{1.90d-10}  & \num{1.13d-11}  & \num{8.53d-12} \\
210  & \num{1.92d-10}  & \num{1.14d-11}  & \num{8.61d-12} \\
220  & \num{1.93d-10}  & \num{1.15d-11}  & \num{8.69d-12} \\
230  & \num{1.95d-10}  & \num{1.17d-11}  & \num{8.76d-12} \\
240  & \num{1.96d-10}  & \num{1.18d-11}  & \num{8.84d-12} \\
250  & \num{1.97d-10}  & \num{1.19d-11}  & \num{8.91d-12} \\
260  & \num{1.99d-10}  & \num{1.20d-11}  & \num{8.98d-12} \\
270  & \num{2.00d-10}  & \num{1.21d-11}  & \num{9.05d-12} \\
280  & \num{2.01d-10}  & \num{1.22d-11}  & \num{9.11d-12} \\
290  & \num{2.02d-10}  & \num{1.23d-11}  & \num{9.18d-12} \\
300  & \num{2.03d-10}  & \num{1.24d-11}  & \num{9.24d-12} \\ \bottomrule
\end{tabular}

\end{table}

\begin{table}[h!]
\caption{Product branching ratios at various temperatures.}
\label{tab:br}
\centering
\begin{tabular}{@{}ccccc@{}}
\toprule
Branching ratios & \ce{CH3 + CH2NH} & \ce{E{-}CH3CHNH + H} & \ce{Z{-}CH3CHNH + H} \\ \midrule
10 K    & 88.6 \%	  & 6.8	\%       & 4.6\%	\\
100 K	& 90.7 \%	  &5.3  \%	     &	4.0  \%  \\
300	K   & 90.4 \%     &5.5	\%       &4.1  \%	\\    \bottomrule
\end{tabular}
\end{table}